\documentclass[aps,prd,twocolumn,showpacs,superscriptaddress,nofootinbib,10pt]{revtex4-2}

\usepackage{acronym}
\usepackage{amsfonts}
\usepackage{amsmath}
\usepackage{amssymb}
\usepackage{bm}
\usepackage{cases}
\usepackage{color}
\usepackage{comment}
\usepackage[dvipsnames]{xcolor}
\usepackage{enumerate}
\usepackage{epsfig}
\usepackage{etoolbox}
\usepackage{fancyref}
\usepackage{fancyhdr}
\usepackage[T1]{fontenc} 
\usepackage{graphicx}
\usepackage{hyperref}
\usepackage{indentfirst}
\usepackage{latexsym}
\usepackage{mathrsfs}
\usepackage{mciteplus}
\usepackage{multirow}
\usepackage{rotating}
\usepackage{scrextend}
\usepackage{soul}
\usepackage{subfigure}
\usepackage{verbatim}
\usepackage{amsthm}

\begin{document}

\title[Black Hole Greybody Factors from Korteweg-de Vries Integrals]{Black Hole Greybody Factors from Korteweg-de Vries Integrals: Computation}

\author{Michele Lenzi}
\email{lenzi@ice.csic.es}
\affiliation{Institut de Ci\`encies de l'Espai (ICE, CSIC), Campus UAB, Carrer de Can Magrans s/n, 08193 Cerdanyola del Vall\`es, Spain}
\affiliation{Institut d'Estudis Espacials de Catalunya (IEEC), Edifici Nexus, Carrer del Gran Capit\`a 2-4, despatx 201, 08034 Barcelona, Spain}

\author{Carlos F. Sopuerta}
\email{carlos.f.sopuerta@csic.es}
\affiliation{Institut de Ci\`encies de l'Espai (ICE, CSIC), Campus UAB, Carrer de Can Magrans s/n, 08193 Cerdanyola del Vall\`es, Spain}
\affiliation{Institut d'Estudis Espacials de Catalunya (IEEC), Edifici Nexus, Carrer del Gran Capit\`a 2-4, despatx 201, 08034 Barcelona, Spain}

\begin{abstract}
It has recently been shown that the dynamics of perturbed non-rotating black holes (BHs) admits an infinite number of symmetries that are generated by the flow of the Korteweg-de Vries (KdV) equation. 
These symmetries lead to an infinite number of conserved quantities that can be obtained as integrals of differential polynomials in the potential appearing in the gauge-invariant master equations describing the BH perturbations, the {\em KdV integrals}.
These conserved quantities are the same for all the possible potentials, which means that they are invariant under Darboux transformations, and they fully determine the BHs transmission amplitudes, or greybody factors, via a moment problem. 
In this paper we introduce a new semi-analytical method to obtain the greybody factors associated with BH scattering processes by solving the moment problem using only the KdV integrals. 
The method is based on the use of Pad\'e approximants and we check it first by comparing with results from the case of a P\"oschl-Teller potential, for which we have analytical expressions for the greybody factors.
Then, we apply it to the case of a Schwarzschild BH and compare with results from computations based on the Wentzel–Kramers–Brillouin (WKB) approximation. 
It turns out that the new method provides accurate results for the BH greybody factors for all frequencies. The method is also computationally very efficient.
\end{abstract}

\maketitle

\section{Introduction}

Black Holes (BHs) are one of the most radical predictions of the General Theory of Relativity~\cite{Einstein:1915by,Einstein:1915ca,Einstein:1916vd}. They represent the most extreme gravitational systems and are defined by the existence of an event horizon that acts as a membrane causally separating the spacetime in the sense that no physical signals can get out from the {\em interior} of the horizon, which constitutes a past-future asymmetry~\cite{Finkelstein:1958zz}. Classical BHs have been widely studied and there are many textbooks describing their main properties and physical consequences~\cite{Hawking:1973uf,Wald:1984cw,Novikov:1989sz,Chandrasekhar:1992bo}. On the other hand, there is accumulating evidence of the existence of astrophysical systems that are compatible with the general relativistic model of a BH. Actually, in some cases the BH model of General Relativity turns out to be the most conservative description in the sense that alternative models require the introduction of exotic forms of matter for which there is no observational evidence. Observations of BH candidates are being made using different types of astronomical techniques, from radio observations to the more recent observations by ground-based gravitational-wave       detectors~\cite{LIGOScientific:2018mvr,LIGOScientific:2021usb,LIGOScientific:2021djp}, including X-rays~(see, e.g.~\cite{2016A&A...587A..61C}). Binary BHs are probably the main source of gravitational waves for current detectors, but also for future ones like third-generation detectors~\cite{Sathyaprakash:2012jk,Evans:2021gyd} and space-based detectors like LISA~\cite{LISA:2017pwj,Amaro-Seoane:2022rxf,LISACosmologyWorkingGroup:2022jok,LISA:2022kgy}.

BHs are becoming central objects for a number of research areas within astrophysics and cosmology, and also in fundamental physics, where progress towards theories that encompass quantum effects and relativistic gravitation use the study of physical phenomena around BHs that goes beyond the current established knowledge. This is partly a consequence of the fact that BHs can have any mass, from the smallest to the largest scales. The only thing we need for their existence is a viable physical mechanism for their formation.

Scattering processes around BHs, together with quasinormal mode oscillations of BHs, are one of main physical processes that can be described using relativistic perturbation theory of BHs (BHPT).  Particles and/or fields scattered by BHs are of great relevance for astrophysics and fundamental physics as they can provide us with key information about the properties of BHs. For this reason, there is a number of studies about BH scattering (see, e.g.~\cite{Vishveshwara:1970zz,Starobinskil:1974nkd,Sanchez:1976fcl,Sanchez:1976xm,Sanchez:1977vz,Futterman:1988ni,Andersson:2000tf,Castro:2013lba,Folacci:2019vtt}), both within classical general relativity and also in semiclassical gravity, where we can gain significant insight into quantum gravitational phenomena.

In the case of non-rotating BHs, the perturbations can be expanded in scalar, vector, and tensor spherical harmonics in such a way that the perturbative equations for each harmonic mode, and for each of the two possible parities, decouple from each other. Moreover, for each mode, the perturbative equations decouple so that the perturbations are governed by a set of master wave-like equations with a potential that tells us the response of the BH to excitations~\cite{Regge:1957td,Cunningham:1978cp,Cunningham:1979px,Cunningham:1980cp,Zerilli:1970la,Zerilli:1970se,Moncrief:1974vm} (see also Ref.~\cite{Chandrasekhar:1992bo}). In the case of scattering processes, the transmission and reflection probabilities can be obtained from the analysis of the master equations. Some low- and high-frequency limits for the greybody factors were obtained in~\cite{Page:1976df,Unruh:1976fm,Starobinsky:1973aij,Starobinskil:1974nkd} and in~\cite{Sanchez:1976fcl,Sanchez:1976xm,Sanchez:1977vz} respectively. Other methods to obtain greybody factors are: Wentzel–Kramers–Brillouin (WKB) approximations~\cite{Schutz:1985km,Iyer:1986np,Iyer:1986nq} (see also~\cite{Konoplya:2003ii,Matyjasek:2017psv,Konoplya:2019hlu}); bounding of Bogoliubov coefficients~\cite{Visser:1998ke,Boonserm:2008zg}; monodromy techniques~\cite{Neitzke:2003mz,Castro:2013lba,Harmark:2007jy}, etc.

Recently~\cite{Lenzi:2022wjv}, we adopted a different point of view to the computation of BH greybody factors, which profits from two previous studies, one on the structure of the space of master functions and master equations~\cite{Lenzi:2021wpc}, and the other one on the symmetries that one can introduce in that space~\cite{Lenzi:2021njy}. The picture that emerged from these studies is that the space of possible master functions and equations is much bigger than what it was previously known, where we can distinguish two different branches of equations. One branch corresponds to master equations with the well-known potentials: the Regge-Wheeler~\cite{Regge:1957td} (odd-parity perturbations) and Zerilli~\cite{Zerilli:1970la} (even-parity perturbations) potentials. The second branch contains an infinite number of master equations with new potentials. In~\cite{Lenzi:2021njy} it was shown that the space of master functions admits different classes of symmetries. On the one hand, all the master equations and functions are connected via Darboux transformations, which shows that all the master functions have the same spectral properties, in particular the same set of quasinormal modes and the same reflection and transmission coefficients. On the other hand, one can also introduce symmetries using the techniques developed for inverse scattering and integrable systems~\cite{Miura:1968JMP.....9.1202M, Gardner:1967wc, Faddeev:1976xar, Deift:1979dt, Novikov:1984id, Ablowitz:1981jq}. We showed~\cite{Lenzi:2021njy} that the master equations admit an infinite number of symmetries generated by the Korteweg-de Vries (KdV) equation~\cite{doi:10.1137/1018076} and the associated infinite hierarchy of non-linear partial differential equations (PDEs). These symmetries translate into conservation laws from which we can obtain an infinite set of conserved quantities~\cite{Miura:1968JMP.....9.1204M, Zakharov:1971faa, Lax:1968fm}, the KdV integrals. Moreover, these symmetries can be seen as isospectral deformations of the master equations that preserve both the scattering transmission coefficient and the quasi-normal modes. Finally, the KdV integrals are the same for all the possible master equations, or in other words, they are invariant under Darboux transformations.

Building on these ingredients, in~\cite{Lenzi:2022wjv} it was shown that the BH greybody factors are completely determined by the KdV integrals associated with the BH potential barrier.  More specifically, the KdV integrals determine, up to a multiplicative factor, the moments of a (probability) distribution function associated with the transmission probability coefficient.  This relationship between moments and distribution is usually known as a {\em moment problem}, the problem of finding the distribution from the moments. It was also argued in~\cite{Lenzi:2022wjv} that the BH moment problem, where the KdV integrals are determined by the BH potential, is determinate in the sense that a solution exists and it is unique. Some other theoretical aspects of this problem in the context of BH perturbations were also discussed. 

~

\noindent{\em Executive Summary}. In this paper, we consider the BH moment problem formulated in~\cite{Lenzi:2022wjv} and develop computational techniques based on Pad\'e approximants to solve the problem.  We first apply these techniques to a case that can be solved analytically, the case of a P\"oschl-Teller potential~\cite{Poschl:1933zz}. Although this potential does not belong to the class of potentials describing BH perturbations, it shares some properties with the BH potential barrier that make it a good touchstone to test methods to describe BH perturbations. We compare our results with exact results of the P\"oschl-Teller potential and show that they coincide to a high degree of approximation. We then apply the technique to the case of the Regge-Wheeler potential and show some error estimation indicators. In this sense, it is important to mention that all the possible BH potentials are physically equivalent in regard to the description of scattering processes (the S matrix), and hence any potential would provide the same results. The error indicators show that our technique to solve the BH moment problem gives accurate results and is very competitive as compared with other techniques, also from the point of view of computational cost.

~

\noindent{\em Structure of the Paper}. In Section~\ref{Sec:master-landscape} we briefly review BHPT for the case of a Schwarzschild BH, introducing the perturbative master equations. In Section~\ref{Sec:kdv-symmetries} we review the important role of KdV symmetries of the master equations describing the BH perturbations and how  the associated KdV integrals uniquely determine the BH greybody factors via a moment problem. In Section~\ref{Sec:pade-method}, we introduce a new method of computing the greybody factors by inverting the BH moment problem using Pad\'e approximants.  In Sec.~\ref{Sec:GF-pade-PT} we discuss the performance of our Pad\'e-based method by comparing results in the case of a P\"oschl-Teller potential~\cite{Poschl:1933zz}, for which we have exact expressions for the greybody factors. In Sec.~\ref{Sec:GF-pade-RW} we apply our Pad\'e-based approximation to perturbations of Schwarzschild BHs, where in all the computations we use the Regge-Wheeler potential, but the results are valid for any potential describing perturbations of a Schwarzschild BH. In Section~\ref{Sec:Conclusions-and-Discussion} we summarize the main results of the paper and discuss further applications of our method and possible avenues for future developments. The paper contains four appendices: In Appendix~\ref{App:Pade-approximants} we introduce some basic elements of Pad\'e approximants; in Appendix~\ref{App:WKB} we quote results on the computation of BH greybody factors using the WKB approximation; in Appendix~\ref{App:PT} we show some results on the P\"oschl-Teller potential; and in Appendix~\ref{App:KdV-PT} we list the first KdV integrals for the P\"oschl-Teller potential.

~

Throughout this paper, otherwise stated, we use geometric units in which $G = c = 1\,$.

\section{Master Equations for BH Perturbations} \label{Sec:master-landscape}

The scattering of test fields, including the gravitational field, and particles by a BH, and also quasinormal oscillations of BHs, are physical processes that can be described by BHPT. In the case of a non-rotating BH, the static background is the Schwarzschild metric
\begin{equation}
d{s}{}^2= {g}^{}_{\mu\nu}dx^{\mu}dx^{\nu}=-f(r)\,dt^2+\frac{dr^2}{f(r)}+r^2d\Omega^2\,,
\label{schwarzschild-metric}
\end{equation}
in which
\begin{equation}
f(r) = 1 - \frac{r^{}_{s}}{r}\,, 
\end{equation}
where $r_s$ is the location of the event horizon, the Schwarzschild radius: $r_{s} = 2GM/c^{2}=2M\,$. Perturbations around a Schwarzschild BH  can be decomposed into scalar, vector, and tensor spherical harmonics thanks to the spherical symmetry of the background. The perturbative Einstein equations decouple for each harmonic $(\ell,m)$ and for each parity~\footnote{A given harmonic component, ${\cal O}^{\ell m}$, is said to be of the even-parity type if, under a parity transformation $(\theta,\phi)$ $\rightarrow$ $(\pi-\theta, \phi+\pi)$, transforms as ${\cal O}^{\ell m} \rightarrow (-1)^{\ell}{\cal O}^{\ell m}$, while it is said to be of the odd-parity type when it transforms as ${\cal O}^{\ell m} \rightarrow (-1)^{\ell+1}{\cal O}^{\ell m}$.}.  The result is a set of ten linear but coupled PDEs for the ten metric perturbations of each harmonic and parity. Gauge choices can help to simplify the problem, but the most important realization is that we can find {\em master functions}, $\Psi^{\rm even/odd}_{\ell m}$, i.e. linear combinations of the metric perturbations and their first-order derivatives, that decouple the system of PDEs for the perturbations. In other words, these master functions satisfy {\em master equations} where no other combinations of perturbations appear. It turns out that these master equations are wave-type equations (see, e.g.~\cite{Martel:2005ir,Nagar:2005ea,Lenzi:2021wpc}):
\begin{equation}
\left(-\frac{\partial^2}{\partial t^2} + \frac{\partial^2}{\partial x^2} - V^{\rm even/odd}_\ell  \right)\Psi^{\rm even/odd}_{\ell m} = 0\,,
\label{master-wave-equation}
\end{equation}
where: $x$ is the {\em tortoise} coordinate, defined by $dx/dr = 1/f\,$; $V^{\rm even/odd}_\ell(r)$ is a potential constructed from the Schwarzschild background and depends on the harmonic number $\ell$ and the parity; the master functions $\Psi^{\rm even/odd}_{\ell m}(t,r)$ depend on both harmonic numbers, the parity and the coordinates of the time-radial sector of the metric, which constitutes a true Lorentzian metric~\footnote{In what follows, for the sake of simplicity, we drop the harmonic numbers $(\ell,m)$ from the master functions.}. 

In~\cite{Lenzi:2021wpc}, all the possible master equations with master functions linear in the metric perturbations and their first-order derivatives were determined.  In this space of possible master functions and equations one can distinguish two branches: The {\it standard branch}, which is characterized by having the known potentials, namely the Regge-Wheeler potential~\cite{Regge:1957td} for odd-parity perturbations
\begin{equation}
V_{\ell}^{\rm RW}(r) = \left(1-\frac{r^{}_s}{r}\right)  \left(\frac{\ell(\ell+1)}{r^{2}} - \frac{3r^{}_{s}}{r^{3}} \right)  \,,
\label{RW-potential}
\end{equation}
and the Zerilli potential~\cite{Zerilli:1970la} for even-parity perturbations 
\begin{eqnarray}
V^{\rm Z}_\ell(r) & = & \frac{f}{\lambda^{2}}\left[ \frac{(\ell-1)^{2}(\ell+2)^{2}}{r^{2}}\left( \ell(\ell+1) + \frac{3r^{}_{s}}{r} \right) 
\right. \nonumber \\
& + & \left. \frac{9r^{2}_{s}}{r^{4}}\left( (\ell-1)(\ell+2) + \frac{r^{}_{s}}{r} \right) \right] \,,
\label{Z-potential}
\end{eqnarray}
where $\lambda$ is a function of $r$ given by
\begin{eqnarray}
\lambda(r) =  (\ell-1)(\ell+2) + \frac{3r^{}_{s}}{r}\,.
\label{lambda-def-sch}
\end{eqnarray}
Regarding master functions, it was found~\cite{Lenzi:2021wpc} that in the odd-parity case the most general master function is a general linear combination of the Regge-Wheeler~\cite{Regge:1957td} and the Cunningham-Price-Moncrief~\cite{Cunningham:1978cp,Cunningham:1979px,Cunningham:1980cp} master functions. For the even-parity sector, it was found that the most general master function is a general linear combination of the Zerilli-Moncrief master function~\cite{Zerilli:1970la,Moncrief:1974vm}, and another master function that appears to be new~\cite{Lenzi:2021wpc}.

The second branch of master functions and equations, named the {\em Darboux branch}, is described in~\cite{Lenzi:2021wpc}. It was shown that it contains an infinite number of different master equations of the form of Eq.~\eqref{master-wave-equation}.  As a consequence, there is an infinite family of possible potentials that have to satisfy a non-linear ordinary differential equation. A remarkable aspect of this branch is that the master functions depend explicitly on the potential. Indeed, for each potential, the master functions are written in terms of the metric perturbations and an integral containing the potential (see~\cite{Lenzi:2021wpc} for the explicit expressions).

The structure of this infinite landscape of master equations and functions was investigated in~\cite{Lenzi:2021njy}. It was shown that all the pairs $(V,\Psi)$ are connected by Darboux transformations (DTs). From a more physical point of view, DTs are isospectral transformations, which means that they preserve the spectrum of the frequency domain operator associated with the master equation, as well as the reflection and transmission coefficients~\cite{Glampedakis:2017rar,Lenzi:2021njy}. Indeed, let us consider single frequency solutions, i.e. let us take the following form of the master function: $\Psi(t,r) = e^{ik t}\psi(x;k)$. Then, the master equation~\eqref{master-wave-equation} becomes a time-independent Schr\"odinger equation of the form
\begin{equation}
\psi^{}_{,xx} - V\psi = -k^2 \psi \,.
\label{schrodinger}
\end{equation}
It was shown in~\cite{Lenzi:2021njy} (see also~\cite{Lenzi:2022wjv}) that DTs map a time-independent Schr\"odinger equation to a physically equivalent one, with different potential barrier and master function, but with the same spectrum.

\section{BH Greybody factors and the Korteweg-de Vries Integrals} 
\label{Sec:kdv-symmetries}

We have mentioned that the different possible master equations describing the first-order perturbations of non-rotating BHs are connected by DTs, which turn out to be isospectral transformations. In addition, there is another type of transformations of the master equations that are isospectral. They consist in deformations~\cite{Lenzi:2021njy,Lenzi:2022wjv} of the time-independent Schr\"odinger equation that follow the Korteweg-de Vries (KdV) equation and the associated infinite hierarchy of KdV equations.  The KdV deformations consist in introducing, in the time-independent Schr\"odinger equation~\eqref{schrodinger}, a dependence on a parameter $\tau$, that is
\begin{equation}
\psi(x)\rightarrow\psi(\tau,x)\,,\;\;
V(x)\rightarrow V(\tau,x)\,,\;\;
k\rightarrow k(\tau)\,,
\end{equation}
in such a way that the potential $V(\tau.x)$ follows the KdV equation
\begin{equation}
V^{}_{,\tau} - 6 VV^{}_{,x} + V^{}_{,xxx}=0 \,,
\label{kdv-equation}
\end{equation}
which is a non-linear PDE.  It is precisely the fact that the deformation of the potential follows the KdV equation that makes the spectrum conserved~\cite{Lenzi:2021njy} (see also~\cite{Lenzi:2022wjv}) under the KdV flow, that is: $(k^2)_{,\tau}=0\,$. This result holds for the continuous spectrum, bound-states and resonances (or quasinormal modes), although in the case of the Schwarzschild BHs, where the potential is positive everywhere, there is no discrete spectrum.

The KdV deformations constitute symmetries of our problem. These symmetries lead to conservation laws which, in turn, lead to conserved quantities~\cite{Miura:1968JMP.....9.1204M, Zakharov:1971faa, Lax:1968fm}, the so-called KdV integrals. In~\cite{Lenzi:2022wjv}, we have showed different paths to these conserved quantities. A particularly interesting approach is the Hamiltonian formulation of the KdV equation, which was first studied by Gardner~\cite{gardner1971korteweg}. Soon afterwards, Zakharov and Faddeev~\cite{Zakharov:1971faa} showed that the results by Gardner and collaborators~\cite{Gardner:1967wc,gardner1971korteweg} can be seen from the point of view of action-angle variables.  The action-angle variables associated to the Hamiltonian formulation of the KdV equation appear naturally when we look at the scattering problem associated with the time-independent Schr\"odinger equation. In this way, Zakharov and Faddeev~\cite{Zakharov:1971faa} were able to show that the KdV equation constitutes a completely integrable Hamiltonian system. Following this line of thought,  we can introduce an infinite hierarchy of KdV evolution equations, associated with the infinite chain of KdV conservation laws, using the Hamiltonian formulation of the KdV equation (see also~\cite{FadeevTakhtajan:1987lda}). The conserved quantities, i.e. the KdV integrals, can be obtained in terms of integrals of differential polynomials of the potential $V$ (that is, polynomials in $V$ and its derivatives).  The explicit expressions of the infinite series of KdV integrals is given by~(see, e.g.~\cite{Zakharov:1971faa}):
\begin{equation}
\mathcal{K}^{}_n  = \int^{\infty}_{-\infty} dx\, \kappa^{}_n(x) \,. 
\label{kdv-integrals}
\end{equation}
where the densities $\kappa_n(x)$ are obtained from the following recurrence relation
\begin{eqnarray}
\kappa^{}_1(x) & = & V(x)\,, \\
\kappa^{}_n(x) & = & - \frac{d}{dx} \kappa^{}_{n-1}(x) - \sum_{k=1}^{n-1} \kappa^{}_{n-k-1}(x) \kappa^{}_k(x)\,, \nonumber \\
&& (n = 2, \ldots ) \,.
\end{eqnarray}

In summary, the potential of a time-independent Schr\"odinger equation~\eqref{schrodinger} has associated an infinite sequence of conserved quantities, the KdV integrals, generated by the KdV flows. We have also seen that BH perturbations admit an infinite number of descriptions in terms of master functions and equations~\cite{Lenzi:2021njy}. The obvious question is how the KdV integrals change for the different master equations (for fixed $\ell$). The answer to this question is very simple: The KdV integrals are the same for all the possible potentials that appear in the master equations describing BH perturbations. In other words, the KdV integrals are invariant under DTs~\cite{Lenzi:2021njy} (see also~\cite{Lenzi:2022wjv} for more details). 

This interplay between BHPT, inverse scattering theory, and integrable systems, in particular the key role of DTs and the KdV equation, is what has motivated the research program initiated in~\cite{Lenzi:2021njy}, and which has led to the result~\cite{Lenzi:2022wjv} that the KdV integrals determine completely the BH greybody factors. This is what has been called the {\it BH moment problem}, that is, the problem of finding the BH greybody factors from the KdV integrals.

BH Greybody factors are defined in the context of BH scattering processes, and they are nothing but the modulus square of the transmission coefficient  through the BH potential barrier. To see this in more detail, let us consider the scattering of a wave coming from $x\rightarrow-\infty$ (the BH horizon) by the BH potential barrier. 
After interacting with the potential barrier, part of the wave is transmitted, goes to $x\rightarrow\infty$, and part is reflected, going back to $x\rightarrow -\infty$. In mathemtical terms, this can expressed in the following way:
\begin{eqnarray}
\label{plane-wave-ab}
\psi(x,k) = \left\{ \begin{array}{lc}
a(k) e^{i k x} + b(k)e^{-i k x}  &  \mbox{for~}x\to -\infty\,,\\[5mm]
e^{i k x}  &  \mbox{for~}x \to \infty\,,
\end{array} \right.
\end{eqnarray}
The coefficients $a(k)$ and $b(k)$ are usually known as the Bogoliubov coefficients. They completely determine the BH scattering matrix as well as the reflection and transmission coefficients as follows
\begin{eqnarray}
t(k) = \frac{1}{a(k)} \,, \quad r(k) = \frac{b(k)}{a(k)} \,.
\label{reflection-transmission-coefficient}
\end{eqnarray}
The transmission and reflection probabilities, or greybody factors, are given by the modulus square of the corresponding coefficients, i.e.
\begin{equation}
T(k)=|t(k)|^2\,, \quad R(k)=|r(k)|^2 \,.
\label{rt-probabilities}
\end{equation}
For real $k$ they satisfy the unitarity condition
\begin{eqnarray}
T(k) + R(k) = 1 \,.
\end{eqnarray}
In Ref.~\cite{Lenzi:2022wjv}, it was shown that the KdV integrals and the BH greybody factors are related by an infinite set of integral relations, the so-called trace identities~\cite{Zakharov:1971faa}, which play an important role in the spectral theory of the time-independent Schr\"odinger equations (see, e.g.~\cite{1989CMaPh.126..379C} and references therein). These identities can be written in the following form
\begin{eqnarray}
\mu^{}_{2j} = \int_{-\infty}^{\infty} dk\, k^{2 j}\, p(k) \,.
\label{moments-hamburger}
\end{eqnarray}
Here, the $\mu_{j}$ are positive constants that coincide with the standard definition of the moments of the (distribution) function $p(k)$. The remarkable fact is that that they are proportional to the KdV integrals as follows
\begin{eqnarray}
\mu^{}_{2j} = (-1)^{j}\,\frac{\mathcal{K}^{}_{2j+1}}{2^{2j+1}} \,,
\label{moments-hamburger-kdv}
\end{eqnarray}
and the distribution function $p(k)$ only depends on the transmission amplitude $T(k)$ 
\begin{equation}
p(k) = - \frac{\ln T(k)}{2\pi} \,,
\label{probability-distribution}
\end{equation}
where the minus sign guarantees that we are expressing everything in terms of positive quantities.
Therefore, the problem of finding the greybody factors of a given potential barrier (without bound states) is equivalent to a \textit{moment problem}~\cite{Shohat1943ThePO,akhiezer1965classical,schmudgen2017moment}. 

The moment problem is a long standing problem in mathematics and it appears in a large variety of physical situations.  Roughly speaking, the moment problem consists in inverting equation~\eqref{moments-hamburger} to find the probability distribution $p(k)$ (only) from the knowledge of the moments $\mu_i$. In our case, the moments are proportional to the KdV integrals associated with the BH potential, and the distribution is a logarithmic function of the greybody factors [see Eqs.~\eqref{moments-hamburger-kdv} and~\eqref{probability-distribution}]. 

To be more precise, the moment problem we are facing is known in the mathematical literature as a \textit{Hamburger} moment problem (see~\cite{Lenzi:2022wjv} for more details), since the interval in which the distribution function $p(k)$ is defined corresponds to the whole real line, and the the moments are computed as integrals over it [see Eq.~\eqref{moments-hamburger-kdv}]. Furthermore, Eq.~\eqref{moments-hamburger} defines a special case of the Hamburger moment problem, namely the {\it symmetric Hamburger moment problem}~\cite{schmudgen2017moment}, as the probability distribution $p(k)$ appears to be a symmetric function of $k$, that is $p(-k) = p(k)$, so that all the odd moments vanish: $\mu_{2j+1}=0\,$. This is an interesting property of the integral equations~\eqref{moments-hamburger} whose nature can be related to the fact that only the odd integrals~\eqref{kdv-integrals} correspond to the true first integrals of the KdV equation (see~\cite{Lenzi:2022wjv} for more details), that is, to the first integrals that arise from the Hamiltonian formulation of the KdV equation. 

The symmetric Hamburger moment problem can be uniquely mapped into an associated {\em Stieltjes moment problem} (there is actually a bijective correspondence~\cite{schmudgen2017moment} between the two moment problems). The Stieljes moment problem has the particularity that the distribution function $p(k)$ is defined in the positive real line. Indeed, we can rewrite Eq.~\eqref{moments-hamburger} in the following form
\begin{eqnarray}
\mu^{}_{2j}
=
2\int_{0}^{\infty}\!\!\! dk\, k^{2 j} p(k)  \,.
\label{moments-stieltjes0}
\end{eqnarray}
We can now introduce the following change of variable:
\begin{equation}
\xi = \sigma^2 k^2 \,,    
\end{equation}
where $\sigma$ is constant with dimensions of length that makes $\xi$ to be a dimensionless quantity. Then, we
can write
\begin{eqnarray}
\mu^{}_{2j}
=  \frac{1}{\sigma^{2j+1}}\int_{0}^{\infty}\!\!\! d\xi\, \xi^{j} \tilde{p}(\xi)
\equiv 
\tilde{\mu}^{}_{j} \,,
\label{moments-stieltjes}
\end{eqnarray}
where  
\begin{equation}
\tilde{p}(\xi) = \frac{p(\xi)}{\sqrt{\xi}} = \frac{p(k)}{\sigma\,k} \,.
\label{stieltjes-to-hamburger}
\end{equation}
The associated Stieltjes distribution has non-vanishing odd and even moments $\tilde{\mu}_{j}$, both defined by the even moments $\mu_{2j}$ of the Hamburger moment problem. We can also introduce dimensionless moments from Eq.~\eqref{moments-stieltjes}  in the following way
\begin{eqnarray}
m^{}_{j} \equiv \int_{0}^{\infty} d\xi\, \xi^{j} \, \tilde{p}(\xi) = \sigma^{2j+1}\tilde{\mu}^{}_{j} \,,
\label{adimensional-moment-problem}
\end{eqnarray}
which provides a formulation of the moment problem in terms of dimensionless quantities only.
Finally, it is worth mentioning that it is possible to show~\cite{schmudgen2017moment} that when the symmetric Hamburger moment problem~\eqref{moments-hamburger-kdv} has a unique solution, so does the associated Stieltjes problem of Eq.~\eqref{moments-stieltjes}.

\section{Solving the Moment Problem using Pad\'e Approximants} \label{Sec:pade-method}

The connection between the scattering trace formulas and the moment problem opens a new avenue to develop novel methods to calculate the transmission (and reflection) coefficients of any potential barrier that does not have bound states (as in the case of positive potential barriers as it happens for non-rotating BHs). In Ref.~\cite{Lenzi:2022wjv}, the moment problem approach has been used to show that the KdV integrals, which are proportional to the moments, completely and uniquely determine the greybody factors of the BH potential barrier. In other words, the BH moment problem associated to the calculation of the BH greybody factors has a unique solution. 

Moreover, in Ref.~\cite{Lenzi:2022wjv} it was described how, when analytical expressions are available, the moment problem can be solved with the use of the Stieltjes-Perron inversion formula~\cite{AFST_1894_1_8_4_J1_0} (see also~\cite{Shohat1943ThePO,akhiezer1965classical,schmudgen2017moment}). This was illustrated by considering the particular case of a P\"oschl-Teller potential~\cite{Poschl:1933zz}. 

However, since it is quite unrealistic to think that we can have analytic expressions for a general potential, as we do for the P\"oschl-Teller case, we present here a general approach to construct approximations to the BH greybody factors.  In this sense, it is important to remark that in the literature there is no a unique general solution method for the moment problem, but various approximate and/or numerical approaches that have been explored during the last century like, for instance: The method of moments and its generalizations~\cite{HULBURT1964555,Marchisio}; kernel density functions approximation~\cite{GAVRILIADIS20124193}; maximum entropy methods~\cite{Meadpapanicolau, TAGLIANI1999291} (see also~\cite{JOHN20072890,lebaz:hal-01876363} for some reviews); etc. 

Since we are interested in exploiting the connection between KdV integrals and greybody factors in a semi-analytical way, we choose to follow the approach of Ref.~\cite{Amindavar1994PadeAO}, where the authors show how to find a probability distribution function starting from the asymptotic form of the corresponding moment generating function (MGF) in combination with the use of Pad\'e approximants (see also Ref.~\cite{baker_graves-morris_1996}). This choice is further motivated by the particular features of the Pad\'e approximants to a Stieltjes series.  The reason is that for Stieltjes series the convergence of diagonal and subdiagonal Pad\'e approximants is guaranteed~\cite{baker_graves-morris_1996, Bender:1978bo}. In what follows we use this method to obtain semi-analytical expressions for the BH greybody factors in terms only of the KdV integrals associated to the BH potential barrier.

The procedure we follow consists in determining the Pad\'e approximants corresponding to the asymptotic series of the MGF and then to take the inverse Laplace transform. We first illustrate how this can be done in practice and then we apply it to the well-known P\"oschl-Teller potential barrier, in Sec.~\ref{Sec:GF-pade-PT}, where we compare our results with exact expressions.  Once we show that the method works properly we apply it, in Sec.~\ref{Sec:GF-pade-RW}, to the BH potential barrier, the potential that appears in the master equation for perturbations of the Schwarzschild metric.

Let us start with the formulation of the Stieltjes moment problem in terms of dimensionless quantities [see Eq.~\eqref{adimensional-moment-problem}].  We have computed analytically a few KdV integrals both for the P\"oschl-Teller potential~\cite{Poschl:1933zz} and also for the Regge-Wheeler potential (and hence for any potential in any of the master equations describing Schwarzschild perturbations).  The ones for the P\"oschl-Teller potential are given in Appendix~\ref{App:KdV-PT} while the ones for the Regge-Wheeler potential are given in Appendix E of Ref.~\cite{Lenzi:2022wjv}.
The constant $\sigma$, in the case of the P\"oschl-Teller potential can be taken to be $\sigma = 1/\alpha$,  while for the Regge-Wheeler potential we can choose it simply as $\sigma = r_s\,$.

The MGF for the moments $m_j$ corresponding to the distribution $\tilde{p}$ is usually defined as the Laplace transform of the (probability) distribution~\cite{billingsley1995probability}
\begin{equation}
M(t) = \int_{0}^{\infty} d\xi\, e^{-t \xi} \, \tilde{p}(\xi)
\equiv
\mathcal{L}(\tilde{p})
\,,
\label{MGF}
\end{equation}
where $\mathcal{L}$ denotes the Laplace transform operator.
By expanding the exponential inside the integral in a Taylor series in $t$, we get the asymptotic behavior of $M(t)$ 
\begin{equation}
 M(t) =
\sum_{n=0}^{\infty} \frac{m_n}{n!}(-t)^n
\,.
\label{MGF-asymptotic}
\end{equation}
Notice that the $n$-th moment can be found as the $n$-th derivative of $M(t)$ at $t=0$, i.e.
\begin{equation}
m_n = (-1)^n M^{(n)}(0)
\,,
\end{equation}
which is the reason why this function is called the MGF. The series in Eq.~\eqref{MGF-asymptotic} is a formal expression that does not necessarily need to converge for our procedure to work~\cite{baker_graves-morris_1996,Bender:1978bo}. 

The step by step procedure to find the probability distribution starting from its moments that we are going to use in this work follows the method proposed in Ref.~\cite{Amindavar1994PadeAO} (see also~\cite{baker_graves-morris_1996}). This method consists in two basic steps: First, to construct the Pad\'e approximants for the asymptotic expansion of the MGF [Eq.~\eqref{MGF-asymptotic}]. And second, to take the Laplace inverse of the Pad\'e approximation to find $\tilde{p}(\xi)$.

As is well known, the Pad\'e approximants are approximations by means of rational functions to a power series, in such a way that the order of accuracy depends on the degree of the polynomials in the numerator and denominator of the rational function (see Appendix~\ref{App:Pade-approximants} for the basic ingredients of Pad\'e approximants that are relevant for this work). Then, the Pad\'e approximants, for the series of $M(t)$ in Eq.~\eqref{MGF-asymptotic}, of order $K+L$, where $K$ and $L$ are positive integers, are written as follows 
\begin{equation}
 \left[K / L \right](t) 
 =
 \frac{P(t)}{Q(t)}
 =
 \sum_{n=0}^{K+L} \frac{m^{}_n}{n!} (-t)^n
 \,.
 \label{pade}
\end{equation}
where $P(t)$ and $Q(t)$ are polynomials of order $K$ and $L$ respectively, that is, 
\begin{equation}
P(t) = \sum_{n=0}^{K} P^{}_n t^n\,, \quad 
Q(t) = \sum_{n=0}^{L} Q^{}_n t^n \,.
\end{equation}
As expected [see Eqs.~\eqref{den-pade} and~\eqref{num-pade} in Appendix~\ref{App:Pade-approximants}], all the coefficients of the polynomials $P_n$ and $Q_n$ are completely determined by the KdV integrals. 
In this work we only consider the diagonal ($K=L$) and sub-diagonal ($K=L-1$) Pad\'e approximants because they play a particular role in the convergence of the Pad\'e sequence as, for example, they provide upper and lower bounds to any Stieltjes series and if the problem is determinate they converge to the unique solution~\cite{Bender:1978bo}.
In order to put ourselves in a situation that would allow us to perform easily the Laplace inversion of Eq.~\eqref{MGF}, we need to first find the poles of the Pad\'e approximants, $t = -t_i\,$. In this way, we can decompose the Pad\'e approximants in partial fractions, that is, we can write them in the form
\begin{equation}
\left[K / L \right](t) 
=
\sum_{i = 1}^{L} \frac{\lambda^{}_i}{t + t_i} \,.
\label{partial-fraction-form-pade}
\end{equation}
Here $K=L,L-1$ and the $\lambda_i$'s are the residues of the poles, i.e.
\begin{equation}
\lambda^{}_i
=
\frac{P(-t^{}_i)}{Q'(-t^{}_i)} \,,
\label{pade-residues}
\end{equation}
where here the prime denotes differentiation with respect to $t\,$. 
The poles of the Pad\'e approximants are expected to be located on the negative half-plane in $t\in\mathbb{C}\,$, therefore we expect $\Re(t_i) > 0\,$~\cite{baker_graves-morris_1996}. Then, as a general rule, whenever we encounter a Pad\'e approximant for which $\Re(t_i) < 0$ for some $i$ (a pole in the positive half-plane), we should discard it as an unacceptable approximation since it clearly has the wrong analytic structure~\cite{baker_graves-morris_1996}. Moreover, since the distribution function is real, we must have either real poles or complex ones coming in pairs together with the complex conjugate. 

The last step is to carry out the inverse Laplace transform of the Pad\'e approximant. This is a simple task since we know that (see, e.g.~\cite{bateman1954tables})
\begin{equation}
\mathcal{L}^{-1}\left( \frac{1}{t + t^{}_i}\right)
=
e^{-t^{}_i\,\xi} \,.
\label{laplace-inversion}
\end{equation}
Therefore, we have the following approximated expression for the distribution function associated with the BH greybody factors
\begin{equation}
\tilde{p}(\xi)
\simeq
\sum_{i = 1}^{L}
\lambda^{}_i \, e^{-t^{}_i\,\xi}  \,,
\label{inverse-laplace-stieltjes}
\end{equation}
from where it is straightforward to obtain an expression for the transmission probability $T(k)$. Indeed, taking into account Eqs.~\eqref{stieltjes-to-hamburger} and~\eqref{inverse-laplace-stieltjes} we get
\begin{equation}
p(k)
\simeq \sigma\,k\;
\sum_{i = 1}^{L} \lambda^{}_i\, e^{-t^{}_i\,\sigma^2 k^2} \,.
\label{inverse-laplace-hamburger}
\end{equation}
Finally, using Eq.~\eqref{probability-distribution} we obtain the following approximate form for $T(k)$
\begin{eqnarray}
T(k) & \simeq & 
\exp{\left(-2\,\pi\, \sigma\,k\; 
\sum_{i = 1}^{L} \lambda^{}_i e^{-t^{}_i\,\sigma^2 k^2}\right)} \nonumber \\
& \equiv & 
T^{}_{[K/L]}(k) \,.
\label{greybody-pade}
\end{eqnarray}
From these expressions it becomes clear that if we allow for poles with positive real part, the solution would be unphysical as it would either describe an infinitely growing greybody factor or decaying one, depending on the sign of the residue.
The expression in Eq.~\eqref{greybody-pade} constitutes a semi-analytic approximation for the transmission probability associated with a potential barrier, a BH barrier in our context, exclusively in terms of their KdV integrals. Indeed, all the coefficients in the approximation are evaluated solely from the KdV integrals~\eqref{kdv-integrals}. To show this explicitly we evaluate the first two Pad\'e approximants. The first is the Pad\'e approximant $[0/1]$:
\begin{equation}
    [0/1](t) = \frac{P^{}_0 }{1 + Q^{}_1\, t} \,,
\end{equation}
It depends only on the first two moments as the coefficients $P_0$ and $Q_1$ are given by
\begin{equation}
P^{}_0 = m^{}_0\,,\quad 
Q^{}_1 = \frac{m^{}_1}{m^{}_0} \,.
\end{equation}
There is a single real pole, which is given by 
\begin{equation}
t^{}_1 = \frac{m^{}_0}{m^{}_1} \,, \quad 
\lambda^{}_1 = \frac{m_0^{2}}{m^{}_1} \,,
\end{equation}
where $\lambda_1$ is the associated residue. Therefore, the greybody factor~\eqref{greybody-pade} has the following simple expression
\begin{equation}
T^{}_{[0/1]}(k) 
= 
\exp{\left(-2\pi \frac{m^{2}_0}{m^{}_1} \sigma\,k\,  e^{- \frac{m^{}_0}{m^{}_1} \sigma^2 k^2}  \right)}
\,.
\end{equation}
The second Pad\'e approximant is $[1/1]$:
\begin{equation}
[1/1](t) = \frac{P^{}_0 + P^{}_1\, t}{1 + Q^{}_1\, t} \,,
\end{equation}
which contains only the first three moments 
\begin{equation}
P^{}_0 = m^{}_0\,,\quad 
P^{}_1 = \frac{m^{}_0 m^{}_2 -2\,m_1^2}{2 m_1}\,,\quad 
Q^{}_1 = \frac{m^{}_2}{2\,m^{}_1} \,.
\end{equation}
Again, there is a single pole which,  together with the corresponding residue, are given by
\begin{equation}
t^{}_1 = \frac{2 m^{}_1}{m^{}_2} \,,\quad 
\lambda^{}_1 = \frac{4 m^{3}_1}{m^{2}_2}
\,,
\end{equation}
and the greybody factor~\eqref{greybody-pade} adopts the following expression
\begin{eqnarray}
T^{}_{[1/1]}(k) 
& = & 
\exp\left(-2\pi \frac{4 m^{3}_1}{m^{2}_2} \sigma\,k\,  e^{-\frac{2 m^{}_1}{m^{}_2} \sigma^2 k^2} \right)
\,.
\end{eqnarray}
These expressions do not provide the most accurate results but may be helpful in cases in which a simple analytical expression is needed.

We can summarize the method we just have shown for constructing approximations for the BH greybody factors in a schematic way as an algorithmic list of steps:
\begin{enumerate}
\item Evaluate the first $n$ KdV integrals for the chosen potential.

\item Obtain the moments from the KdV integrals and construct the MGF by using the expansion~\eqref{MGF-asymptotic} at order $n$.

\item Construct Pad\'e approximants of order $[K/L]$, with $K+L \leq n$.

\item Evaluate the poles and residues [see Eq.~\eqref{pade-residues}] of the Pad\'e approximants and discard those with poles that have positive real part.

\item Apply the Laplace inversion formula~\eqref{laplace-inversion} to finally obtain the approximations [see Eq.~\eqref{greybody-pade}] for the greybody factors.
\end{enumerate}

\section{Greybody Factors for the P\"oschl-Teller Potential} \label{Sec:GF-pade-PT}

In order to test our method for the computation of greybody factors, it is convenient to compare our results with a case in which we have analytical expressions for them. In the case of BH perturbations we do not have exactly solvable potentials, and for this reason we consider the well-known case of the P\"oschl-Teller potential~\cite{Poschl:1933zz}:
\begin{eqnarray}
V^{}_{\rm PT}(x) =
\frac{U^{}_0}{\cosh^2{(\alpha x)}}
=
\frac{\alpha^2 (\beta^2+\frac{1}{4})}{\cosh^2(\alpha x)} \,,
\label{pt-potential-barrier}
\end{eqnarray}
where $\beta$ is a dimensionless constant given by 
\begin{equation}
\beta = \sqrt{\frac{U^{}_0}{\alpha^2} - \frac{1}{4}} > 0 \,.
\end{equation}
This is one of the few cases of a potential, defined on the whole real line, in which the time-independent Schr\"odinger equation~\eqref{schrodinger} is exactly solvable~\cite{landau1981quantum} (see Appendix~\ref{App:PT} for a summary of the main relevant results).
The greybody factors for the P\"oschl-Teller potential are given by Eq.~\eqref{pt-transmission-real-k}, i.e.
\begin{eqnarray}
T^{}_{\rm PT} =
\frac{\sinh^2\left(\frac{\pi k}{\alpha}\right)}{\cosh^2\left(\frac{\pi k}{\alpha}\right)+\sinh^2\left(\pi \beta\right)}
\,.
\label{pt-transmission-real-k-2}
\end{eqnarray}
Exact solvability, together with the similarity in shape\footnote{We can adjust the P\"oschl-Teller potential parameters to have a potential barrier that looks quite similar to the Schwarzschild BH potential barrier. Nevertheless, both potentials differ in the decay to zero near spatial infinity, which has important consequences for the time-dependent response to perturbations: In the BH case the long-term behavior is dominated by power law tails~\cite{Leaver:1986gd} in contrast with the P\"oschl-Teller case, where there are no tails.} with the BH potential barrier, makes it a perfect playground to get deeper insight on the BH scattering (see e.g. Ref.~\cite{Ferrari:1984zz,Ferrari:1984ozr}). In this section, we exploit this comparison to test our approximation based on Pad\'e approximants.

\begin{figure}[t]
\centering
\includegraphics[width=8cm,height=6.5cm]{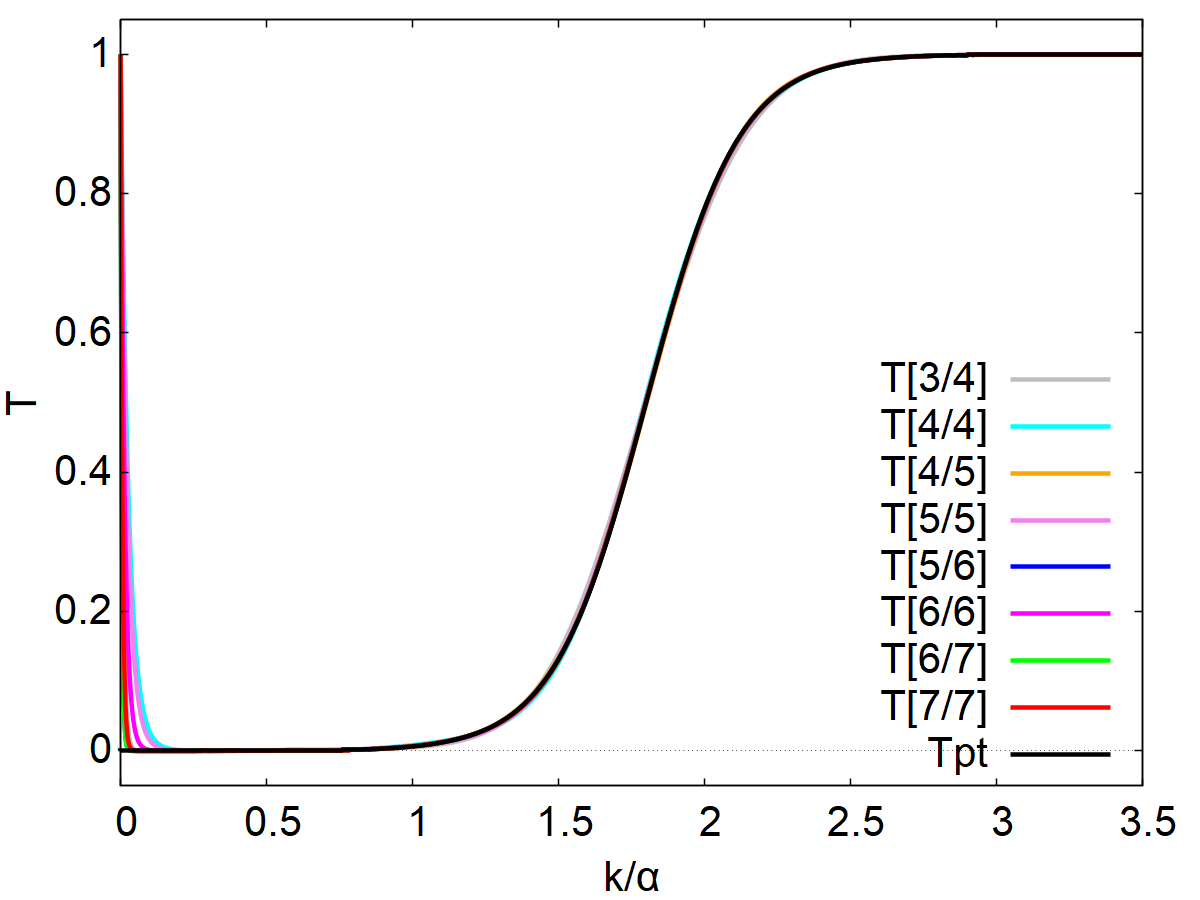}
$\ $
\includegraphics[width=8cm,height=6.5cm]{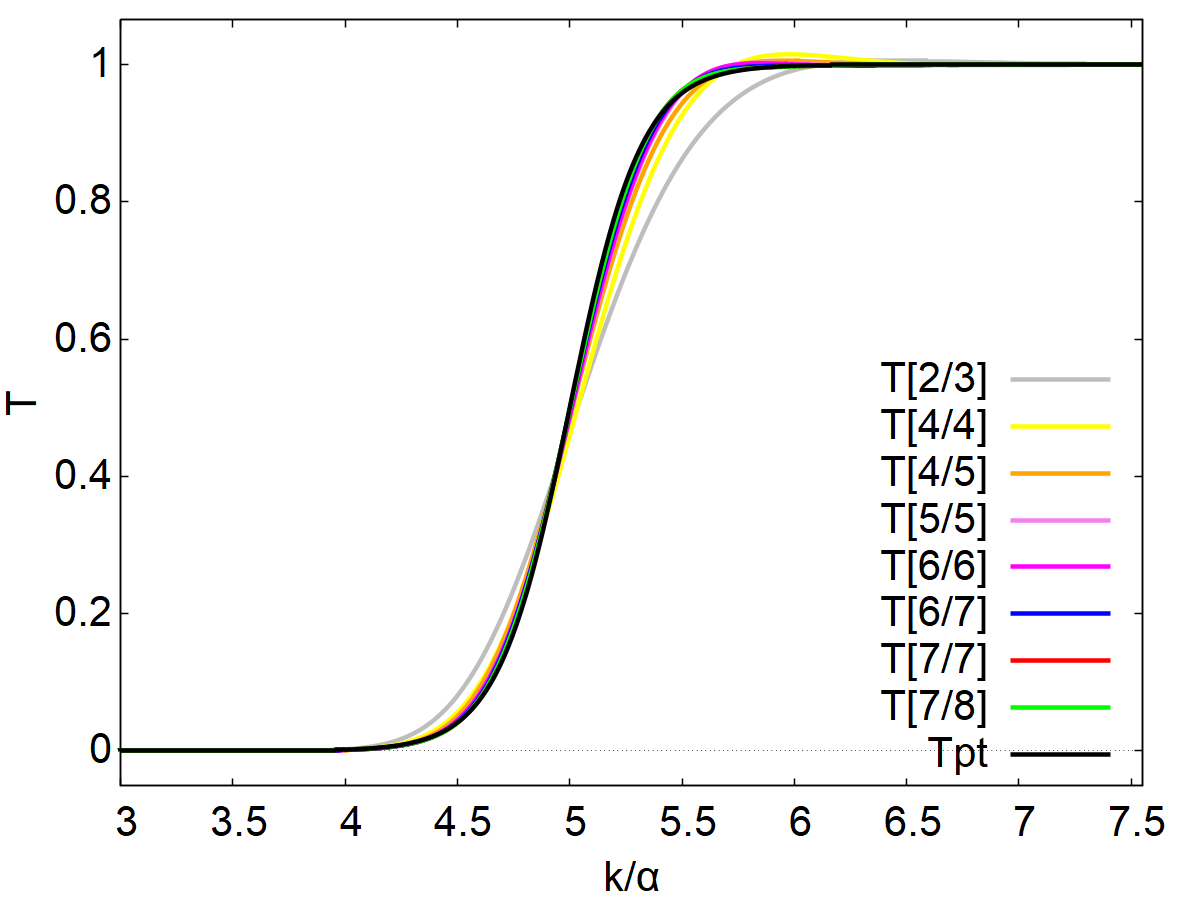}
\caption{Greybody factors for the P\"oschl-Teller potential barrier~\cite{Poschl:1933zz} with respect to the wave number $k$ obtained from the Pad\'e approximation of Eq.~\eqref{greybody-pade} and compared with the exact solution in Eq.~\eqref{pt-transmission-real-k-2} for $\beta=1.8$ (top panel) and $\beta=5$ (bottom panel). The notation $T[N/M]$ means that the greybody factor is computed from the $[N/M]$ Pad\'e approximant to the MGF asymptotic series. 
\label{Tpade-Tpt}}
\end{figure}

Let us now apply the method we described in Sec.~\ref{Sec:pade-method} (the steps involved are outlined at the end of the section) to find approximations for the greybody factors of the P\"oschl-Teller potential barrier.  We start with the computation of the KdV integrals. The first ten non-zero integrals for the the P\"oschl-Teller potential are given in Appendix~\ref{App:KdV-PT}. From these integrals, we find the moments, construct the Pad\'e approximants together with their poles and residues, and finally apply the inverse Laplace transform. If we consider the first six non-zero KdV integrals, the $T_{[2/3]}(k)$ approximation to the transmission coefficient is given by (we set the P\"oschl-Teller parameter $\beta = 5$):
\begin{eqnarray}
T^{}_{[2/3]} & \simeq &  \exp \left\{ - \frac{2\pi k}{\alpha}\left[ 14.895\, e^{-0.256 \frac{k^2}{\alpha^2}}\sin\left(0.093\, \frac{k^2}{\alpha^2}\right) \right. \right. \nonumber
\\
&-& 5.586\, e^{-0.256\, \frac{k^2}{\alpha^2}} \cos{\left( 0.093\, \frac{k^2}{\alpha^2}\right) } \nonumber
\\
&+& \left. \left. 19.167\, { e^{-0.743\, \frac{k^2}{\alpha^2}}} \right] \right\} \,,
\end{eqnarray}
where the trigonometric functions appear due to the presence of pairs of complex conjugate poles. A similar structure is shared by the other  $T_{[K/L]}$ functions.  Different comparisons of the approximations obtained with our method with the exact solution for $T(k)$ [see Eq.~\eqref{pt-transmission-real-k-2} and Appendix~\ref{App:PT}] are given in Fig.~\ref{Tpade-Tpt} for different values of the parameter $\beta$. As we can see, the approximation~\eqref{greybody-pade} proves to be highly accurate, even when we use low-order Pad\'e approximants. Indeed, we can see in Fig.~\ref{Tpade-Tpt} that the approximations based on the first Pad\'e approximants are almost indistinguishable from the exact solution.

\begin{figure}[t]
\centering
\includegraphics[width=8cm,height=6.5cm]{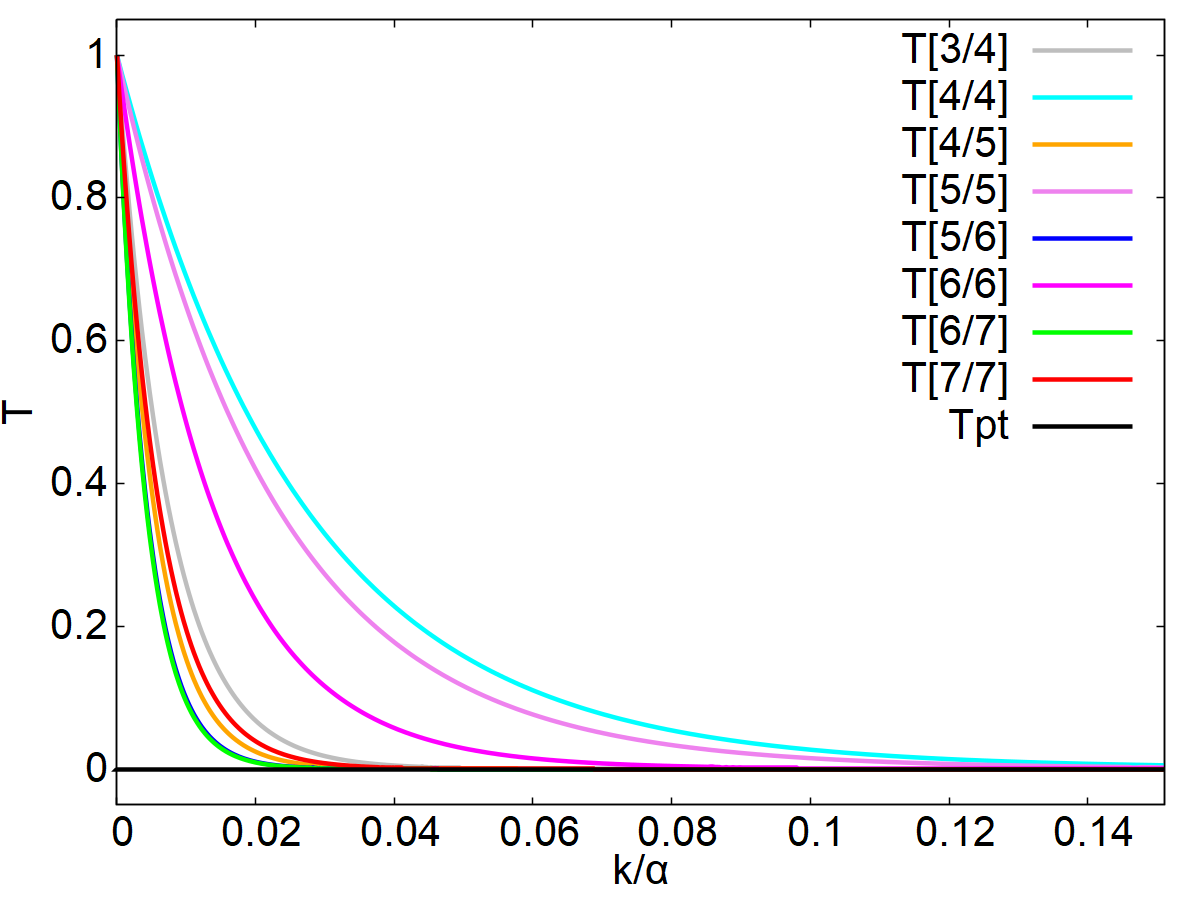}
$\ $
\includegraphics[width=8cm,height=6.5cm]{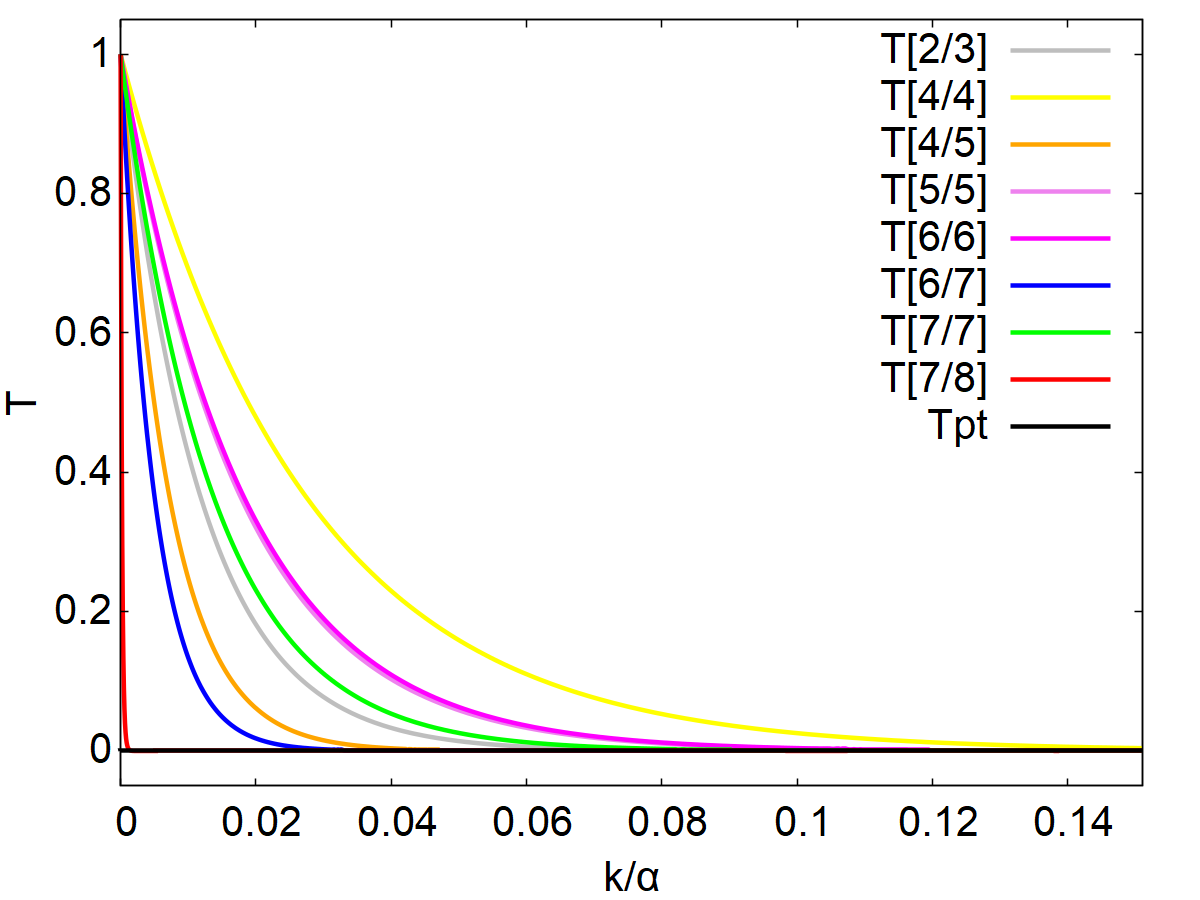}
\caption{Plots of the behavior of the greybody factors for the P\"oschl-Teller potential in the low-frequency (low $k$) regions obtained by zooming near $k=0$ in the plots of Fig.~\ref{Tpade-Tpt} for both $\beta=1.8$ (top panel) and $\beta=5$ (bottom panel). \label{Tpade-Tpt-zoom}}
\end{figure}

The main drawback of the semi-analytic approximation represented by Eq.~\eqref{greybody-pade} is the not so good behavior at small frequencies (small $k$). However, as one can see in Fig.~\ref{Tpade-Tpt-zoom}, the region in which the behavior deviates from the expected one is pushed towards the origin as we include more KdV integrals in the approximation. Or in other words, the region were the approximation gets worse is shrinked towards the origin as we use high-order Pad\'e approximants. This suggests that the approximation given by Eq.~\eqref{greybody-pade} converges also at low frequencies in the limit in which we include infinite moments (an infinite number of KdV integrals).

\begin{figure}[t]
\centering
\includegraphics[width=8cm,height=6.5cm]{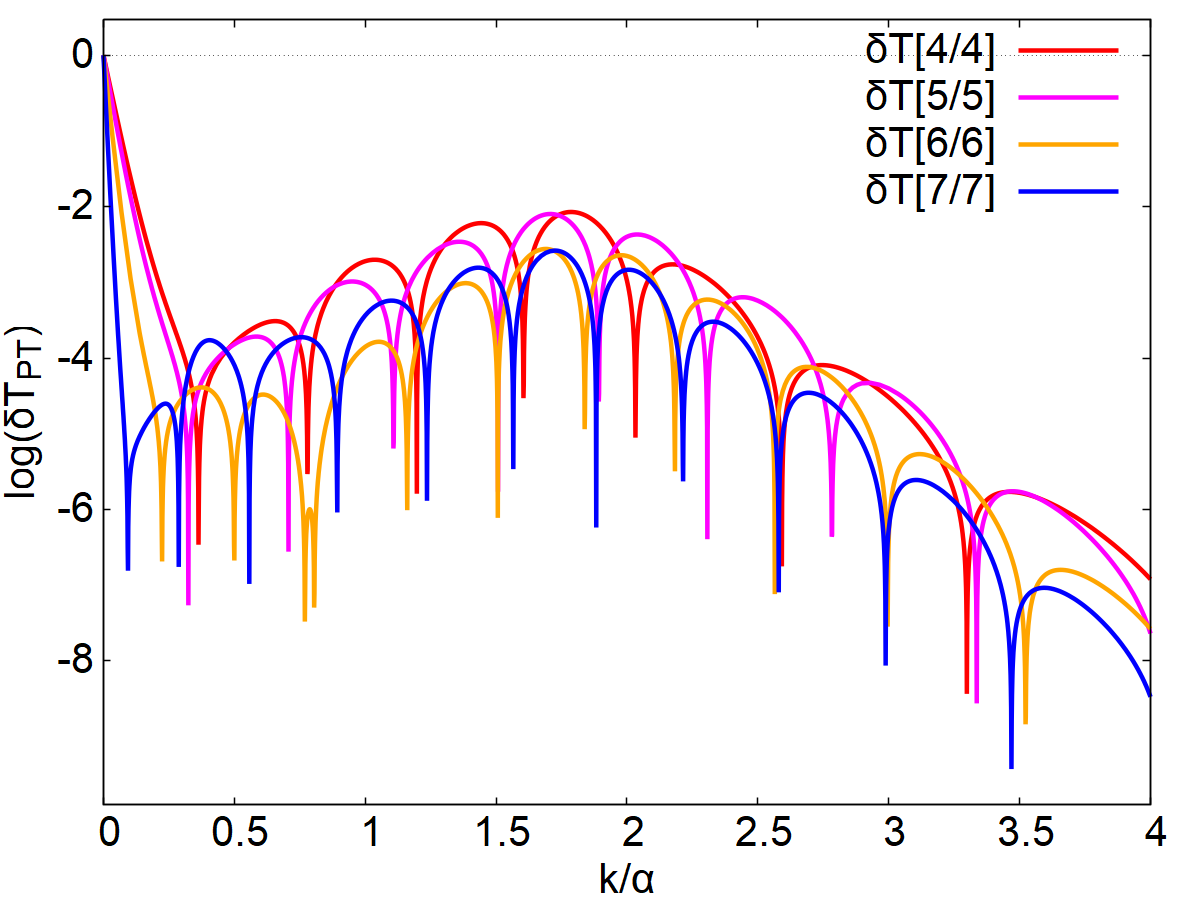}
$\ $
\includegraphics[width=8cm,height=6.5cm]{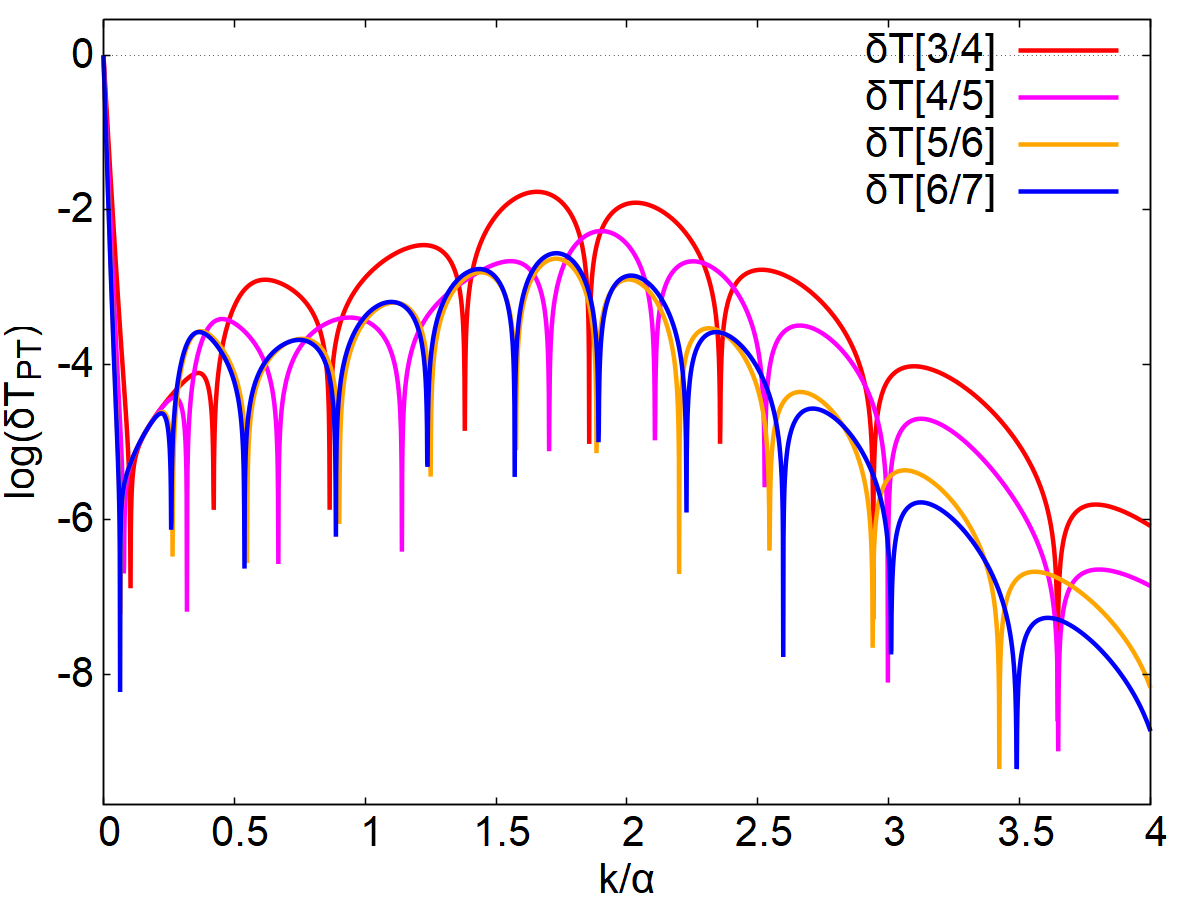}
\caption{Plots of the error estimate $\delta T_{\rm PT}(k)$ [see Eq.~\eqref{delta-Tpt}], in logarithmic scale of base 10, for the diagonal (top panel) and sub-diagonal (bottom panel) Pad\'e-based approximation of Eq.~\eqref{greybody-pade} for the greybody factors with $\beta=1.8\,$. 
\label{deltaTpt-med}}
\end{figure}

\begin{figure}[t]
\centering
\includegraphics[width=8cm,height=6.5cm]{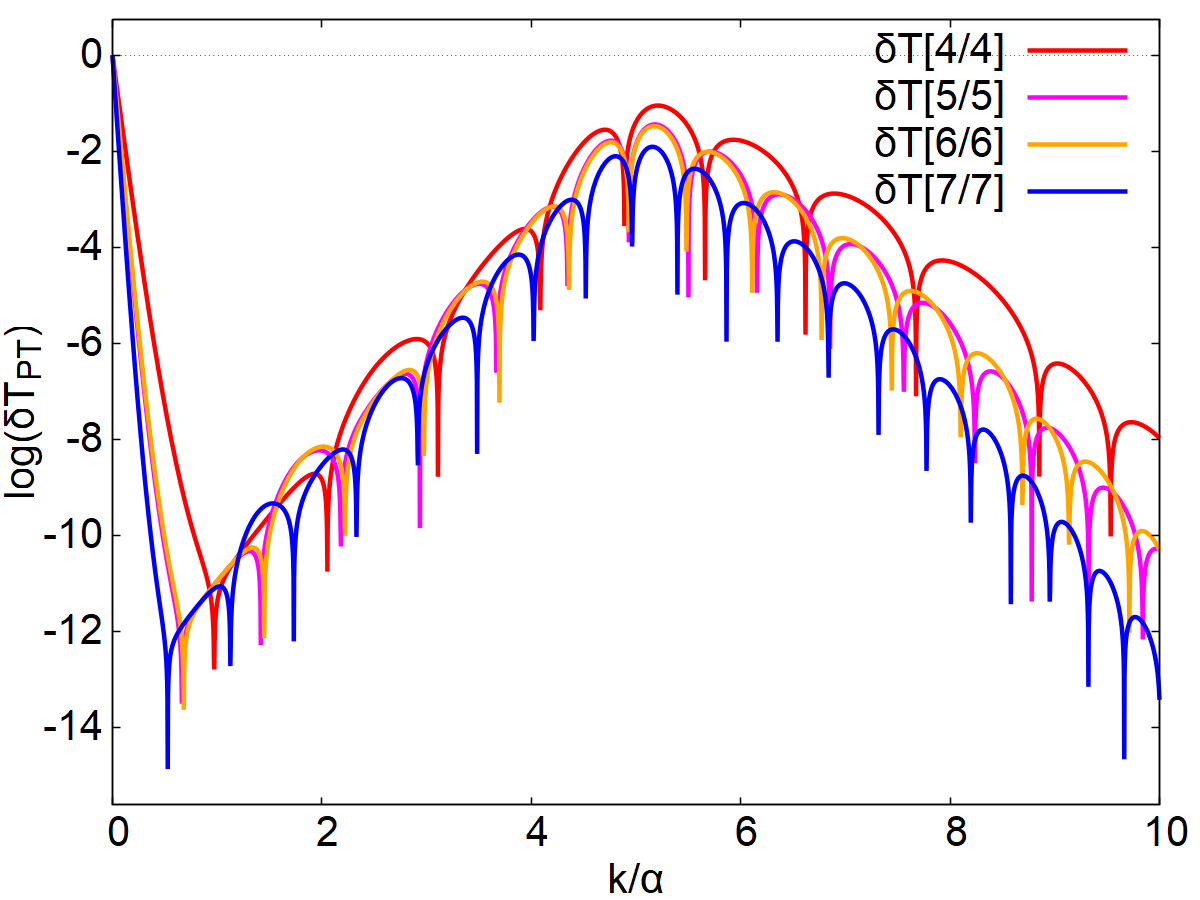}
$\ $
\includegraphics[width=8cm,height=6.5cm]{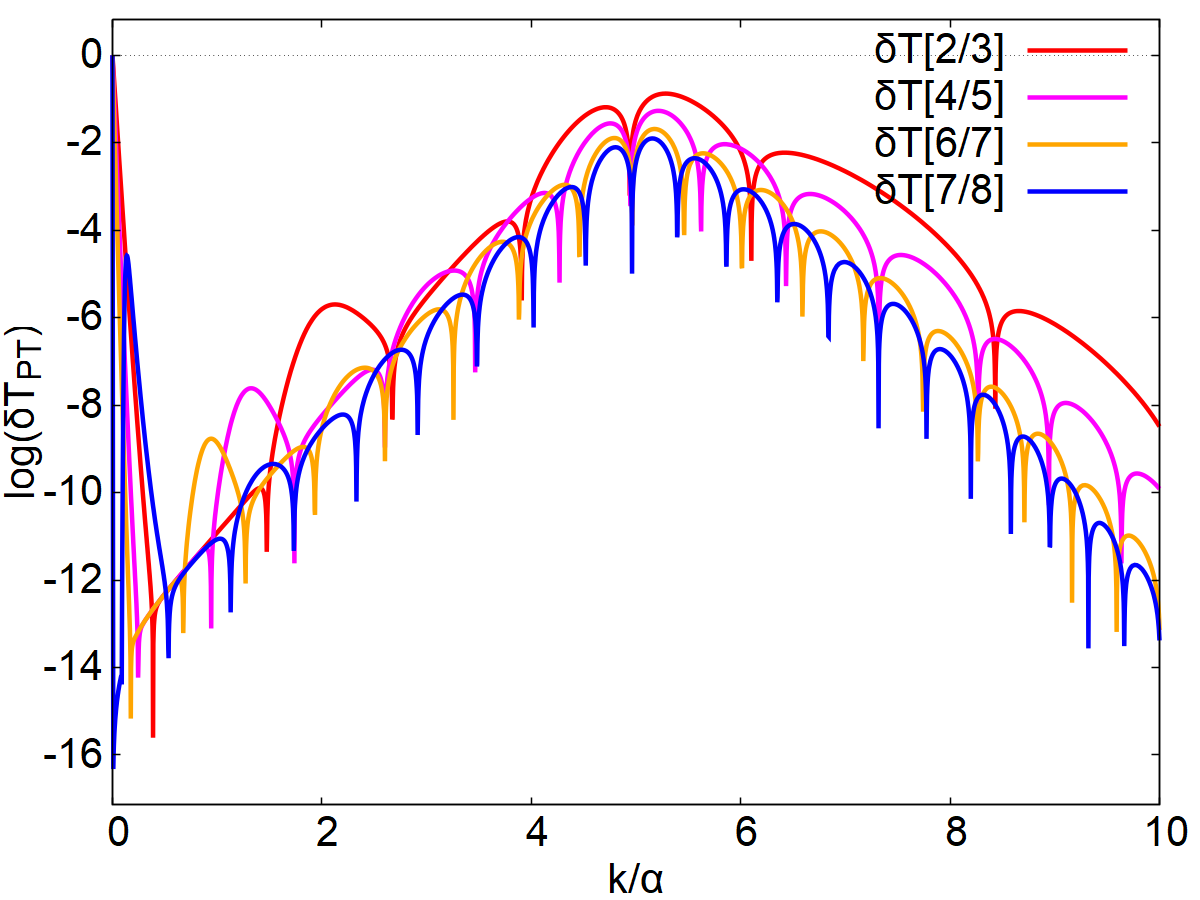}
\caption{Plots of the error estimate $\delta T_{\rm PT}(k)$ [see Eq.~\eqref{delta-Tpt}], in logarithmic scale of base 10, for the diagonal (top panel) and sub-diagonal (bottom panel) Pad\'e-based approximation of Eq.~\eqref{greybody-pade} for the greybody factors with $\beta=5\,$. 
\label{deltaTpt-high}}
\end{figure}

In order to obtain a better quantitative understanding of the behavior of our approximation technique, we need to find appropriate error estimates.  The fact that we are using the P\"oschl-Teller potential to test our method allows us to make direct comparisons with the exact greybody factor in Eq.~\eqref{pt-transmission-real-k-2} and have a better insight of the convergence properties of the procedure. In this sense, let us consider the following $k$-dependent error function
\begin{equation}
\delta T^{}_{\rm PT}(k)
=
\left| T^{}_{\rm PT}(k)  -  T^{}_{[K/L]}(k) \right| \,,
\label{delta-Tpt}
\end{equation}
that is, the absolute difference between the exact solution and our approximation.
Plots of this error estimate are shown in Figs.~\ref{deltaTpt-med} and~\ref{deltaTpt-high} for different values of the parameter $\beta$. As we can see from these plots, both diagonal and subdiagonal Pad\'e approximants to the MGF seem to produce a converging approximation to the exact value. Apart from the region near the origin $k=0$, the error estimate in the intermediate region appears to be slightly worse than in the rest of the $k$ line. Nevertheless, we still obtain a  very good degree of precision in that intermediate region, and this improves with the number of KdV integrals that we include in the approximation. 

Regarding the  error coming from the low-frequency behavior (low $k$), the results shown in Figs.~\ref{deltaTpt-med} and~\ref{deltaTpt-high} clarify that the deviations become significant only in a small area close to the origin $k=0$.  The size of this region decreases as we increase the order of the approximation (or the number of KdV integrals used), so that we can say that we obtain an accurate description also at small frequencies (small $k$). 

The comparison between Figs.~\ref{deltaTpt-med} and~\ref{deltaTpt-high} further shows that the higher the potential barrier is the better is the approximation of Eq.~\eqref{greybody-pade} to the exact value $T_{\rm PT}$, both at low and high frequencies, while it provides slightly worse results in the intermediate region (see also Figs.~\ref{Tpade-Tpt} and~\ref{Tpade-Tpt-zoom}).
On the other hand, it is worth mentioning that due to the shape of $T_{\rm PT}(k)$ and $T_{[K/L]}(k)$, the relative error would not provide a fully meaningful estimate. Indeed, since $T_{\rm PT}$ goes to zero when $k$ does, the contributions from small frequencies to the relative error grow near $k=0$ and hence, we would obtain a misleading measure of the error with respect to the exact solution.

\begin{figure}[t]
\centering
\includegraphics[width=8cm,height=6.5cm]{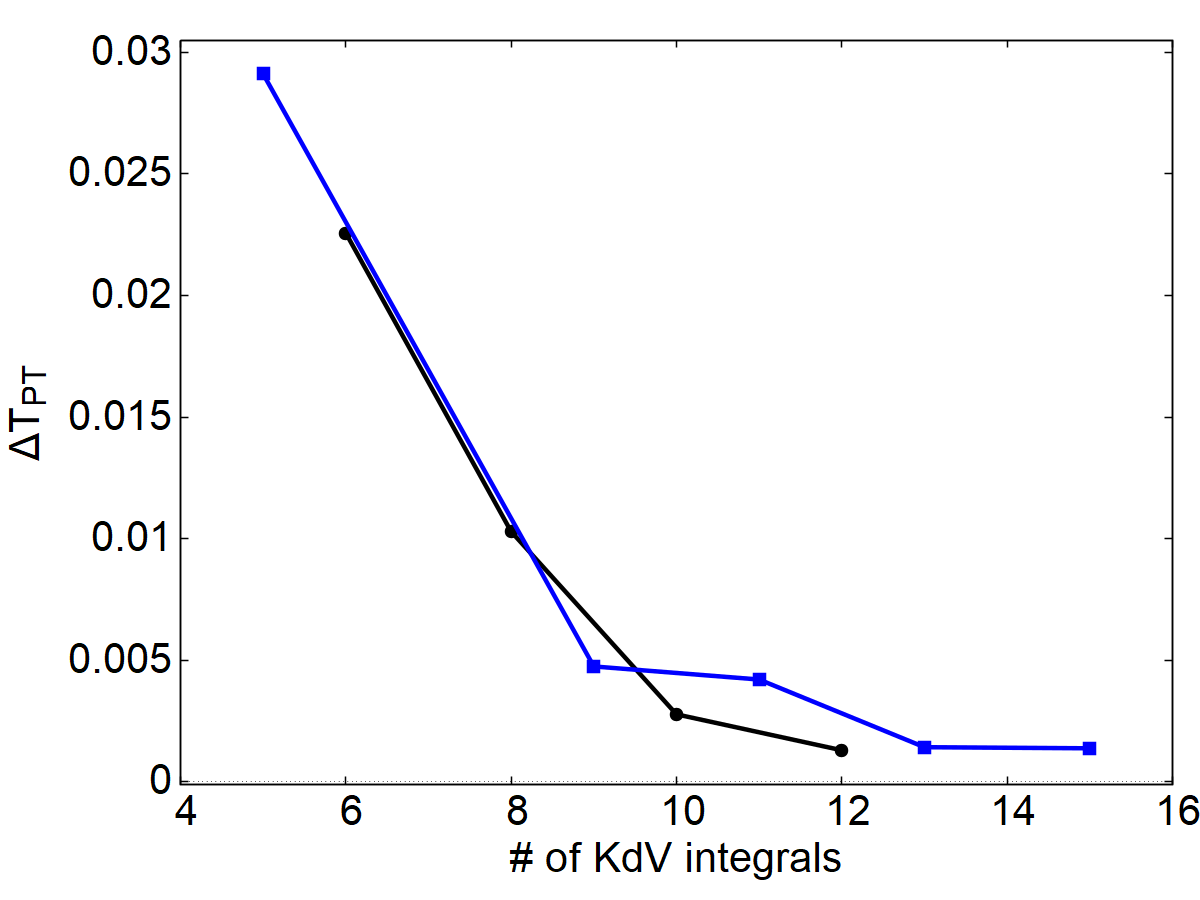}
$\ $
\includegraphics[width=8cm,height=6.5cm]{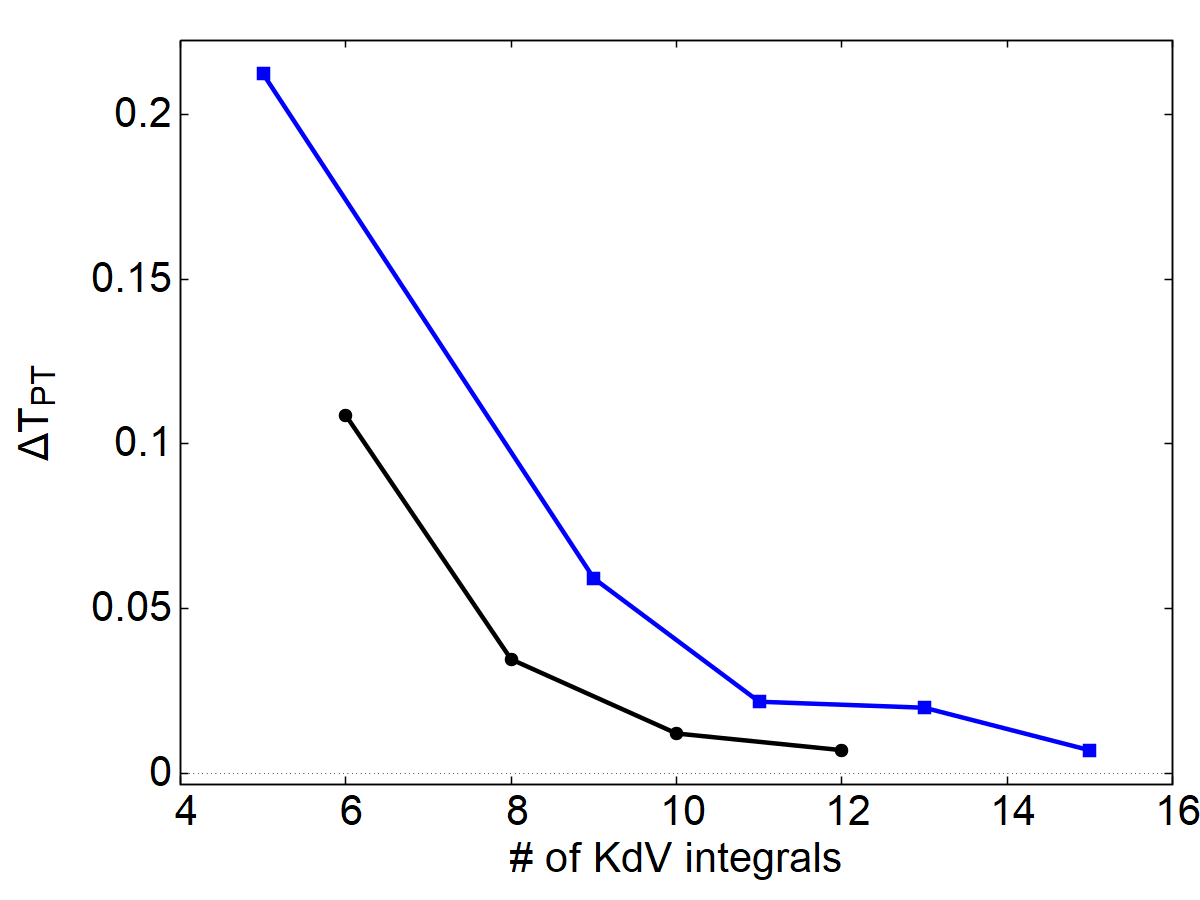}
\caption{Plots of the error $\Delta T_{\rm PT}$ [Eq.~\eqref{real-error}] as computed from diagonal (blue line) and sub-diagonal (black line) Pad\'e approximations with cut-offs $(k_0\,,\,k_{\infty}) = (0.25\,,\,20)$ and with P\"oschl-Teller parameter $\beta=1.8$ (top panel) and $\beta=5$ (bottom panel).  
\label{DELTA-Tpt}}
\end{figure}

The integration of $\delta T_{\rm PT}(k)$ over $k$ provides a global error indicator that can also inform us about the improvement of the approximation. Then, we define
\begin{equation}
\Delta T^{}_{\rm PT} 
=
\int_{k^{}_0}^{k^{}_{\infty}} dk \, \delta T^{}_{\rm PT}(k)\,,
\label{real-error}
\end{equation}
where $k_0$ is a cut-off at low frequencies and $k_{\infty}$ at high frequencies. The low-frequency cut-off $k_0$ is introduced to separate the very low-frequency contributions, where our approximations may misbehave, but without changing the global qualitative behavior. The high-frequency cut-off $k_{\infty}$ is introduced instead for computational convenience, since the greybody factors become almost constant for large $k$ and hence, integrating all the way to $k\rightarrow \infty$ may produce meaningless results from the numerical point of view.  In Fig.~\ref{DELTA-Tpt} we show results of the computation of $\Delta T_{\rm PT}$, where we can see that, irrespective of the details of the approximate greybody factors [see Eq.~\eqref{greybody-pade}] at each range of frequencies, the convergence of both diagonal and sub-diagonal Pad\'e approximants is guaranteed.

\begin{figure}[t]
\centering
\includegraphics[width=8cm,height=6.5cm]{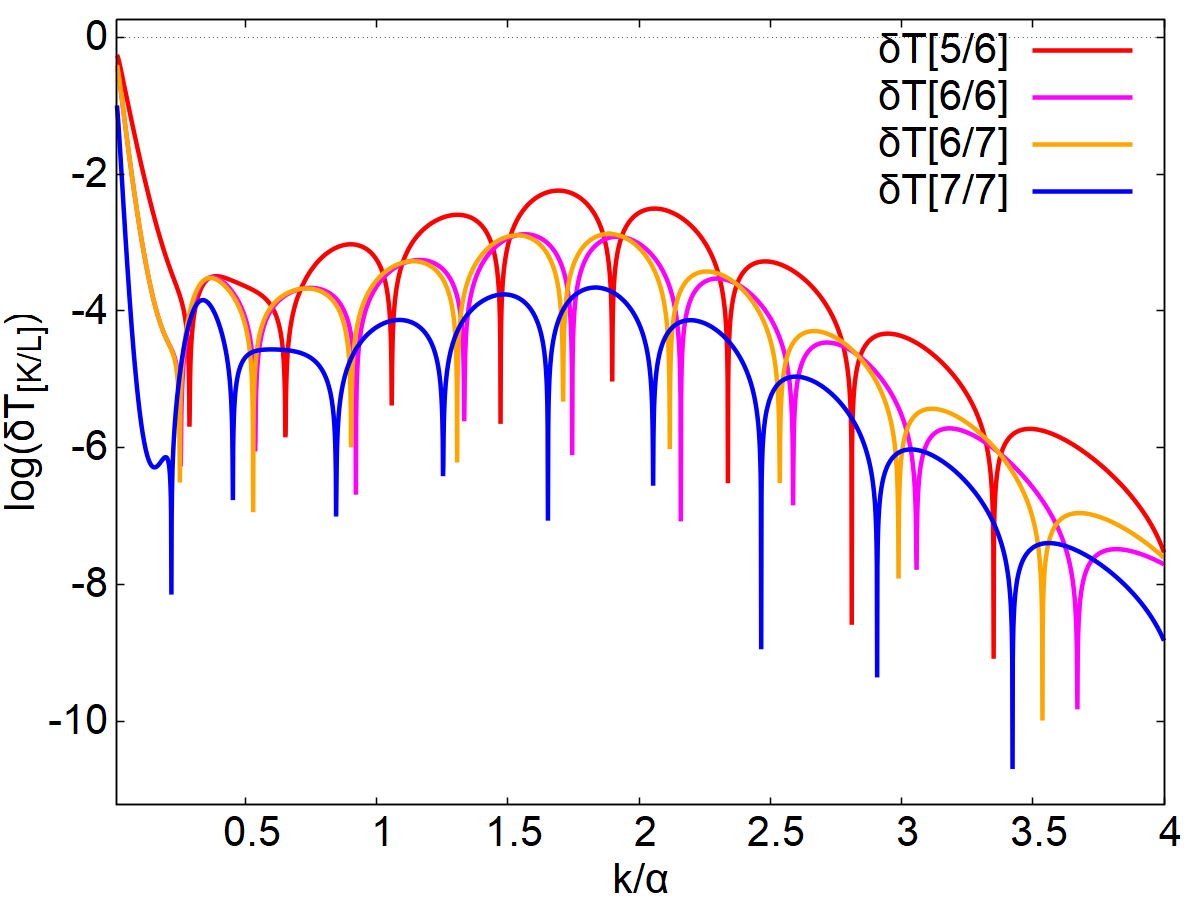}
$\ $
\includegraphics[width=8cm,height=6.5cm]{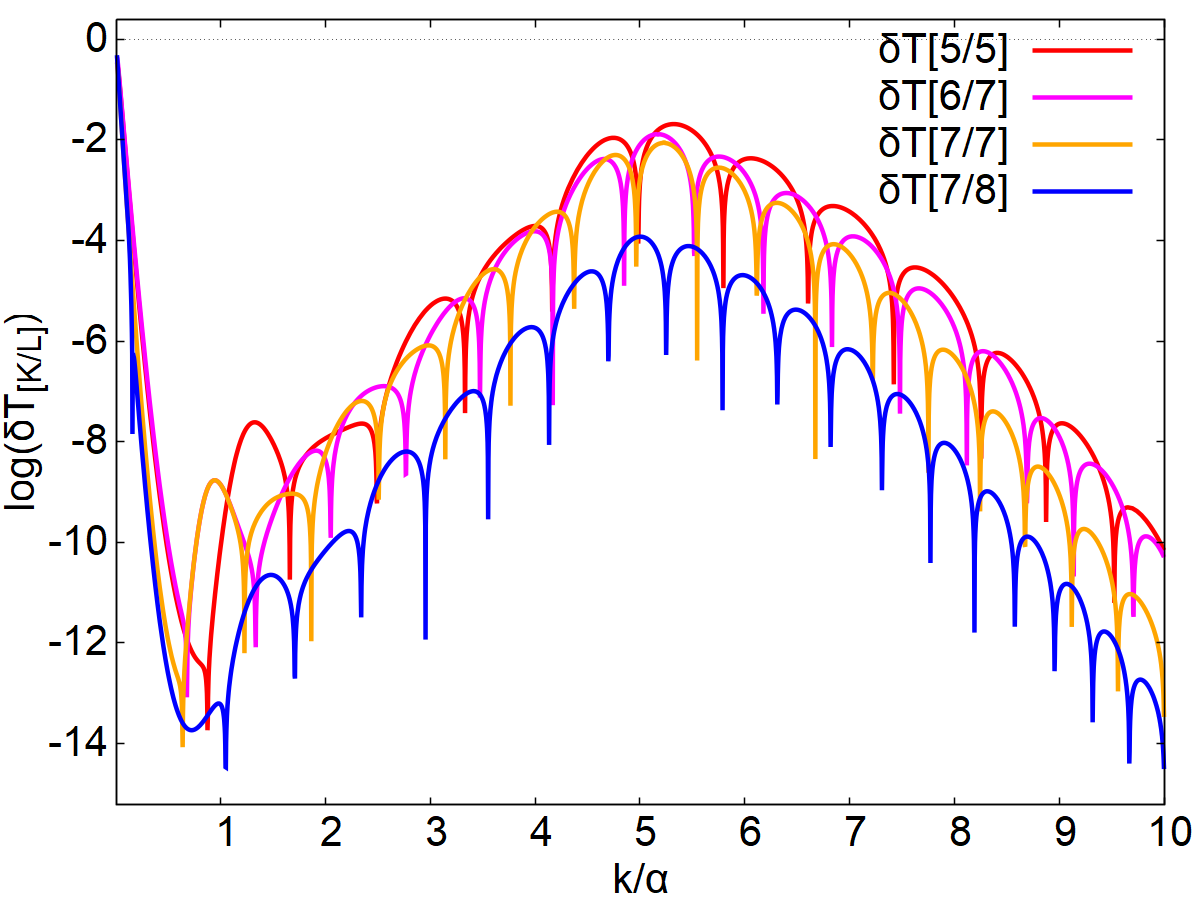}
\caption{Plots of the error estimate in Eq.~\eqref{abs-error-T} for successive approximations to the greybody factor of the P\"oschl-Teller potential, in logarithmic scale of base 10, for $\beta= 1.8$ (top panel) and for $\beta =5$ (bottom panel).
\label{deltaTsucc-PT}}
\end{figure}

The diagonal and subdiagonal Pad\'e approximants provide two different kinds of approximations for the greybody factors in terms of convergence of our Pad\'e-based approximation to the real function~\eqref{pt-transmission-real-k-2}. Each of them follow its own pattern of convergence. This can clearly be seen when studying Stieltjes sequences~\cite{baker_graves-morris_1996, Bender:1978bo}, as already mentioned in Sec.~\ref{Sec:pade-method}. However, for our purposes, it can be useful to study the improvements in the approximation when considering the addition of just one more KdV integral. This turns out to be crucial when the exact solution is not provided, as in the relevant case of a BH potential barrier (see Sec.~\ref{Sec:GF-pade-RW}). To that end, we can introduce an error estimator for successive approximations as follows
\begin{equation}
\delta T^{}_{[K/L]}(k)
=
\left| T^{}_{[(K+d)/K]}(k)  -  T^{}_{[K/L]}(k) \right|
\,,
\label{succ-error-T}
\end{equation}
where either $L=K$ and $d=-1$ or $L=K+1$ and $d=0$.
This is a measure of how much the Pad\'e-based approximation of Eq.~\eqref{greybody-pade} improves every time that we include in the analysis one more moment/KdV integral. The results are shown in  Fig.~\ref{deltaTsucc-PT}, where we can see how the convergence of the approximation with $n+1$ moments improves up to more than one order of magnitude with respect to the one calculated from $n$ moments.

\begin{figure}[t]
\centering
\includegraphics[width=8cm,height=6.5cm]{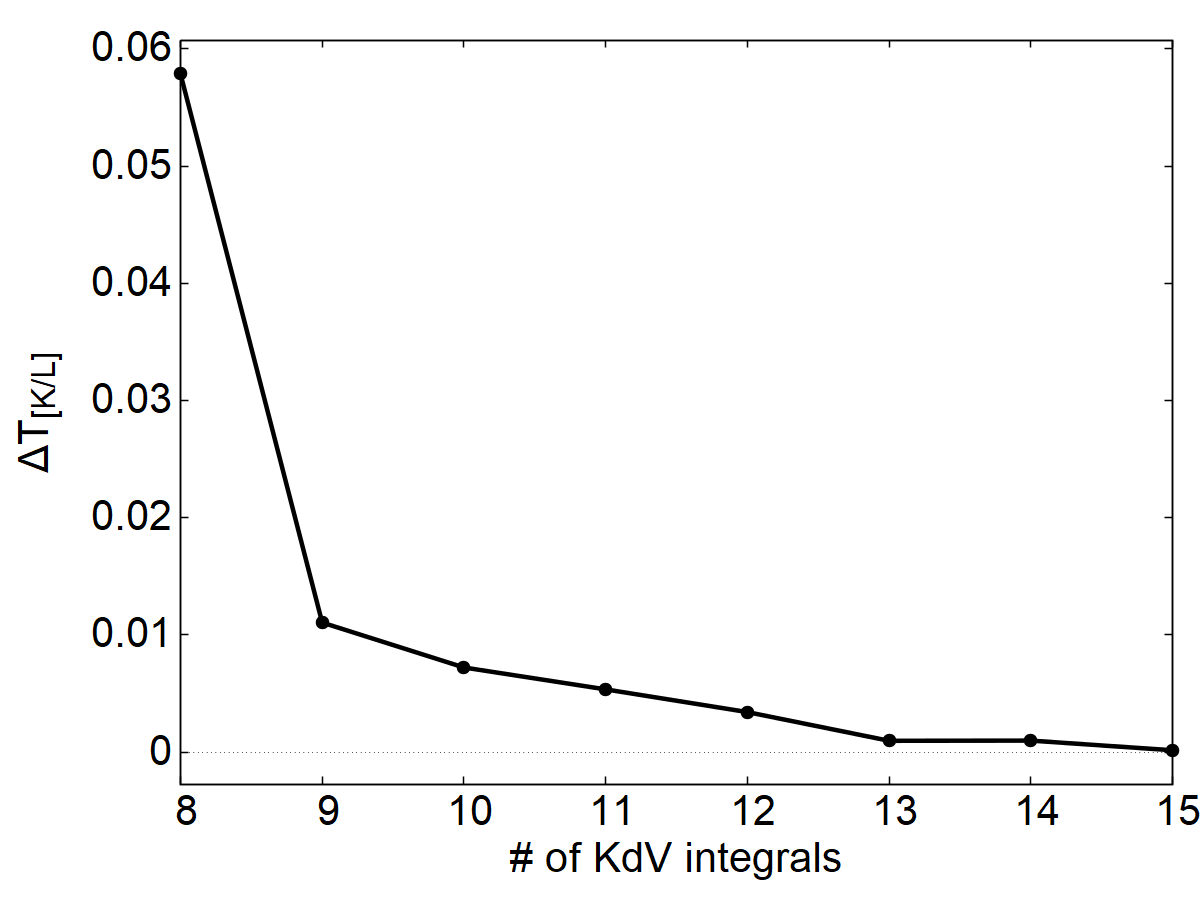}
$\ $
\includegraphics[width=8cm,height=6.5cm]{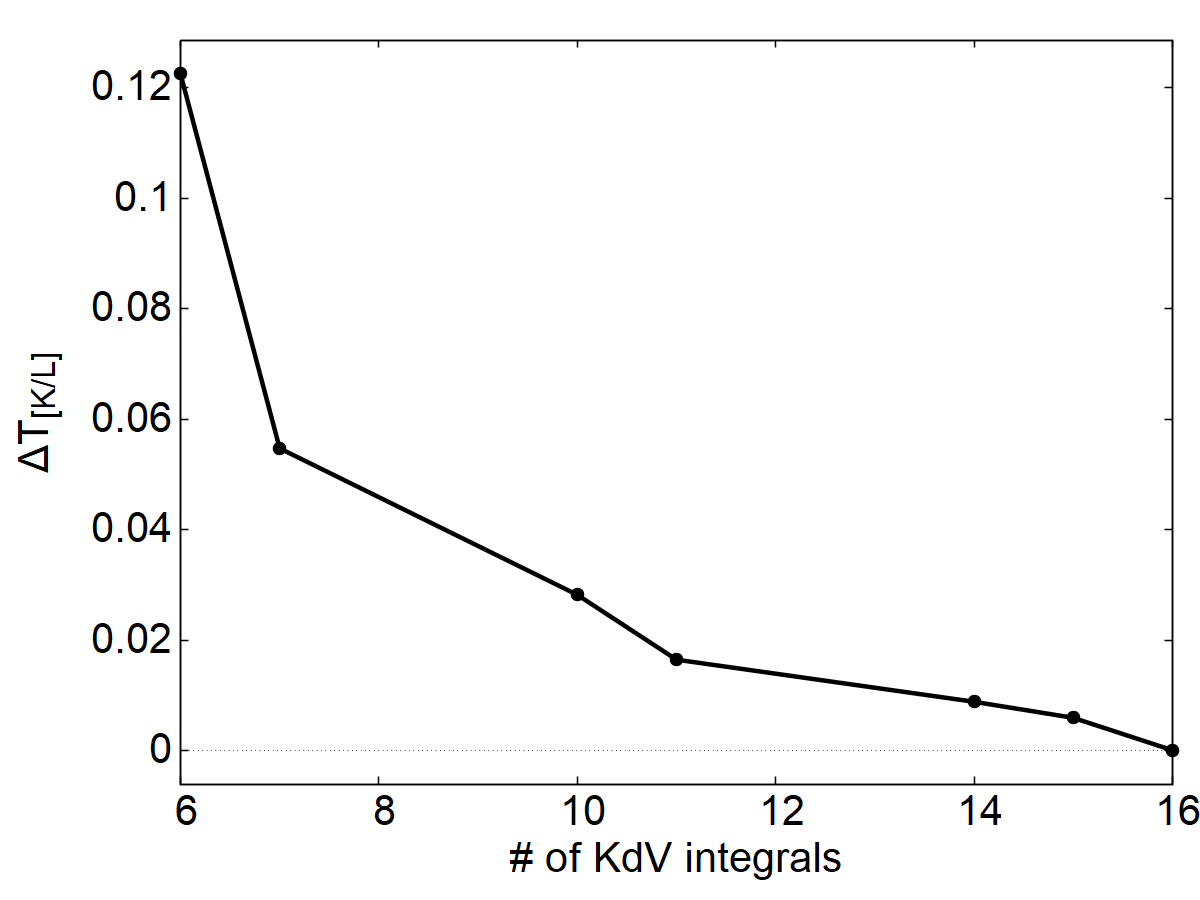}
\caption{Plots for the {\it global} error estimator $\Delta T_{[K/L]}$ [see Eq.~\eqref{abs-error-T}] for successive Pad\'e-based approximations of the P\"oschl-Teller gredybody factors with increasing number of moments/KdV integrals, with cut-offs $(k_0\,,\,k_\infty) = (0.25\,,\,20)\,$, for $\beta=1.8$ (top panel) and $\beta=5$ (bottom panel). 
\label{Delta-error-succ}}
\end{figure}

We can introduce a global error estimate associated with the error estimate in Eq.~\eqref{succ-error-T} for succesive Pad\'e-base approximations [see Eq.~\eqref{greybody-pade}] by integrating this error function over $k$: 
\begin{equation}
\Delta T^{}_{[K/L]} 
=
\int_{k^{}_0}^{k^{}_{\infty}} dk \,\delta T^{}_{[K/L]}(k)\,,
\label{abs-error-T}
\end{equation}
where we have introduced again the cut-offs $(k_0,k_\infty)$ for the same reasons described above. This global error estimate is shown in Fig.~\ref{Delta-error-succ}.
By looking at Figs.~\ref{deltaTsucc-PT} and~\ref{Delta-error-succ} we see that $\delta T_{[K/L]}(k)$ and $\Delta T_{[K/L]}$ have the same qualitative behavior as $\delta T_{\rm PT} (k)$ and $\Delta T_{\rm PT}$, thus the same conclusions can be drawn here. There error estimates are going to be particularly useful in Sec.~\ref{Sec:GF-pade-RW}, where we deal with the Regge-Wheeler potential.

It is also worth mentioning that from these results it appears that the cases in which we have Pad\'e approximants that present positive poles, these poles  do not completely spoil the convergence. Instead, they just appear to slow down the convergence rate of the approximation.

\begin{figure}[t]
\centering
\includegraphics[width=8cm,height=6.5cm]{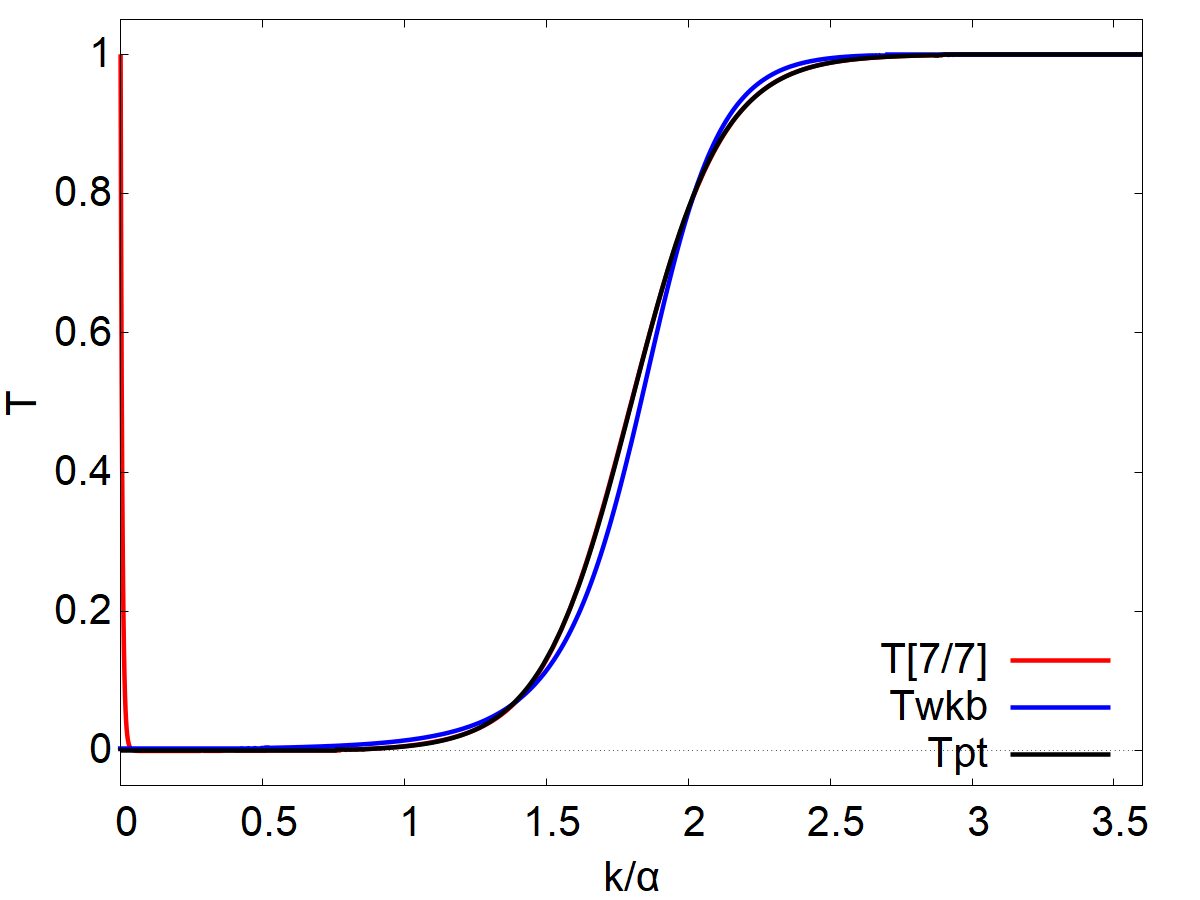}
$\ $
\includegraphics[width=8cm,height=6.5cm]{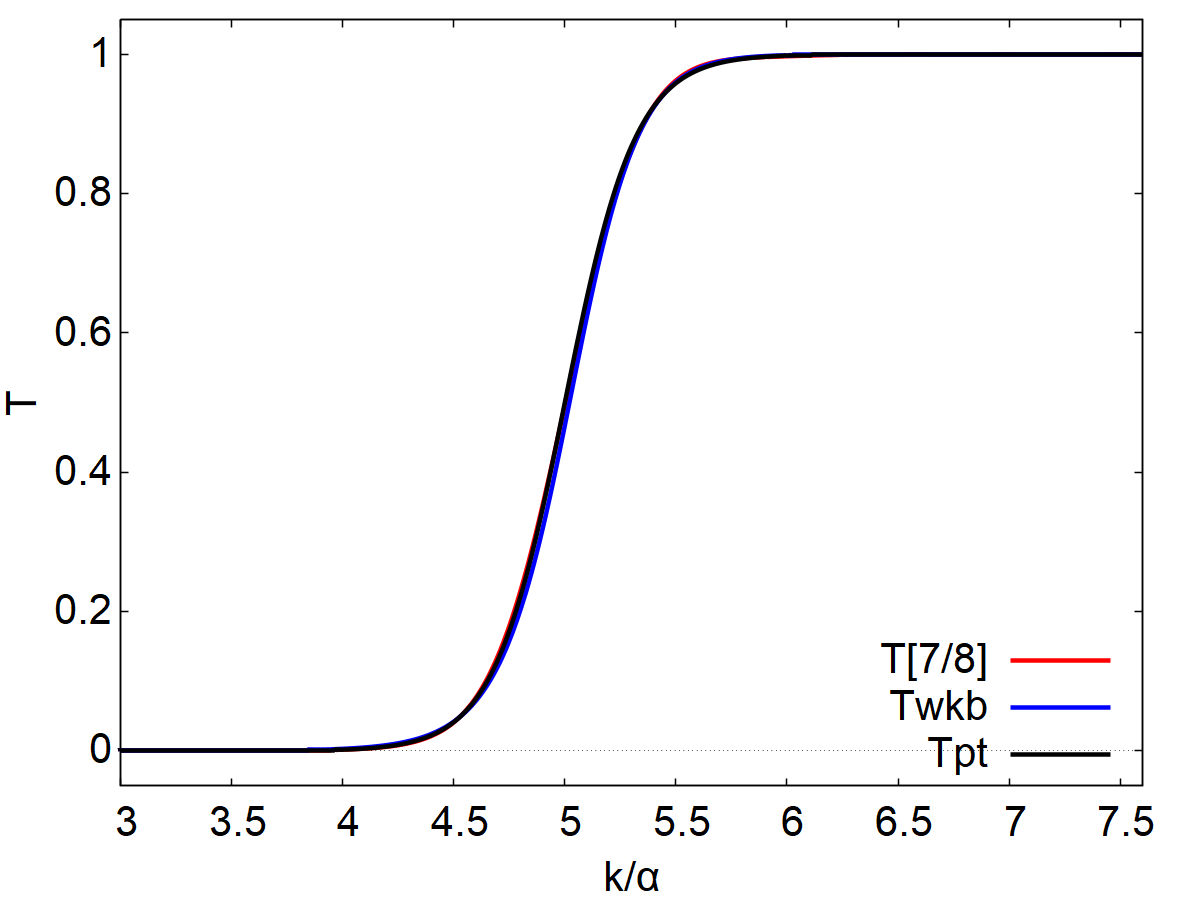}
\caption{Plots of the comparison of computations of the greybody factors for the P\"oschl-Teller potential using the Pad\'e-baed approximation of Eq.~\eqref{greybody-pade}, the WKB approximation method of Eq.~\eqref{greybody-WKB}, and the exact values in Eq.~\eqref{pt-transmission-real-k} for $\beta=1.8$ (top panel) and $\beta=5$ (bottom panel).
\label{Tpade-Tpt-Twkb}}
\end{figure}

In addition to the comparisons with the exact expressions that we have for the P\"oschl-Teller potential and between successive approximations, there is a different type of test of our approximation method that we can carry out, namely the comparison with WKB approximations for the computations of greybody factors~\cite{Schutz:1985km,Iyer:1986np,Iyer:1986nq,Konoplya:2019hlu}.  The expressions of the WKB approximations that we use in this work are shown in Appendix~\ref{App:WKB}.  The results of the comparisons with third-order WKB approximations (the fourth order happens to vanish for the P\"oschl-Teller potential) are shown in Fig.~\ref{Tpade-Tpt-Twkb}. It is clear from these plots that the accuracy of the WKB approximation depends on the parameter $\beta$, which controls the strength of the P\"oschl-Teller potential. Indeed, it is well known that the WKB approximation works better for higher values of the angular momentum in the case of Regge-Wheeler potential~\cite{Schutz:1985km,Iyer:1986np,Iyer:1986nq,Konoplya:2019hlu}. Since the higher the angular momentum is the higher the maximum of the potential barrier becomes, we can apply the same argument to the  case of the P\"oschl-Teller potential in terms of the parameter $\beta$. 
In this sense, our Pad\'e-based approximation seems to be less sensitive to the potential strength. 

On the other hand, our approach differs from the WKB approximation in the way that we build more accurate estimations of the greybody factors. In the case of the WKB method, it is an accumulative process in which we compute higher orders of an approximation series. The size of the expression for each order of approximation in the WKB method grows considerably as we can see in the Appendix~\ref{App:WKB}. In our approach, we can start from an arbitrary number of KdV integrals (moments) and compute the desired order of approximation from the associated Pad\'e approximant. The computational cost of the KdV integrals is quite low as we can see from their explicit expressions in Appendix~\ref{App:KdV-PT} (see also Appendix E in Ref.~\cite{Lenzi:2022wjv}). Moreover, the computational cost of our Pad\'e-based approximation is quite affordable since the evaluation of high-order Pad\'e approximants only require a few seconds using a computer algebra system like Maxima~\cite{maxima} on a personal laptop computer.

Finally, it is interesting to note that both approximations, our Pad\'e-based approximation and the WKB approximation, fail in a similar way in describing the intermediate frequency region. In fact, in Fig.~\ref{Tpade-Tpt} we see that, for high barriers (in the case of the figure $\beta=5$), our Pad\'e -based approximation, from low to high frequencies, grows up to a maximum slightly greater than one, in the intermediate region, before settling down to a constant value equal to one. The same happens in the case of the WKB approximation in the low barrier case as one can see by looking closely at Fig.~\ref{Tpade-Tpt-Twkb} (in the case of the figure $\beta=1.8$), and it becomes more evident for lower barriers (see also Figs.~\ref{Tpade-schw} and~\ref{Tpade-Twkb-schw} for the Regge-Wheeler case in the next section.

\section{Greybody Factors for Schwarzschild Black Holes} \label{Sec:GF-pade-RW}

The greybody factors associated with gravitational perturbations around a Schwarzschild BH background can be evaluated in an analogous way as we have done in the previous section for the case of a P\"oschl-Teller potential~\cite{Poschl:1933zz}. We just need to follow the step-by-step procedure outlined at the end of Sec.~\ref{Sec:pade-method}. There is however a significant difference between the two potentials, beyond exact solvability, and it has to do with the decay towards infinity of the potentials. Indeed, while both potentials decay exponentially as we go towards $x\rightarrow -\infty$ (the BH horizon in the case of the BH potential barrier), the approach towards $x\rightarrow \infty$ (spatial infinity in the case of the BH potential barrier) is very different. In the case of the P\"oschl-Teller potential, it also decays exponentially, while in the case of the BH potential barrier, take the Regge-Wheeler potential as an example, it decays quadratically, that is $V_{\rm RW}\sim L/r^2 \sim L/x^2$, where $L=\ell(\ell+1)$. As we can see, the so-called angular momentum barrier dominates at $x\rightarrow \infty$ in the BH caes. The different decay towards $x\rightarrow\infty$ has important consequences for the properties of the solutions of the time-independent Schr\"odinger equation~\eqref{schrodinger}, and even more in the time-domain. Indeed, it turns out that in the case of a non-rotating BH, the structure of the Green function present a branch cut that is responsible for the power-law tails that we observe in the time-evolution of BH perturbations (see~\cite{Leaver:1986gd} for more details), and which are not present in the case of the P\"oschl-Teller potential in Eq.~\eqref{pt-potential-barrier}. Nevertheless, for the computation of the KdV integrals that we need for the greybody factors, both potentials decay sufficiently fast so that the corresponding integrals are well defined (finite).

\begin{figure}[t]
\centering
\includegraphics[width=8cm,height=6.5cm]{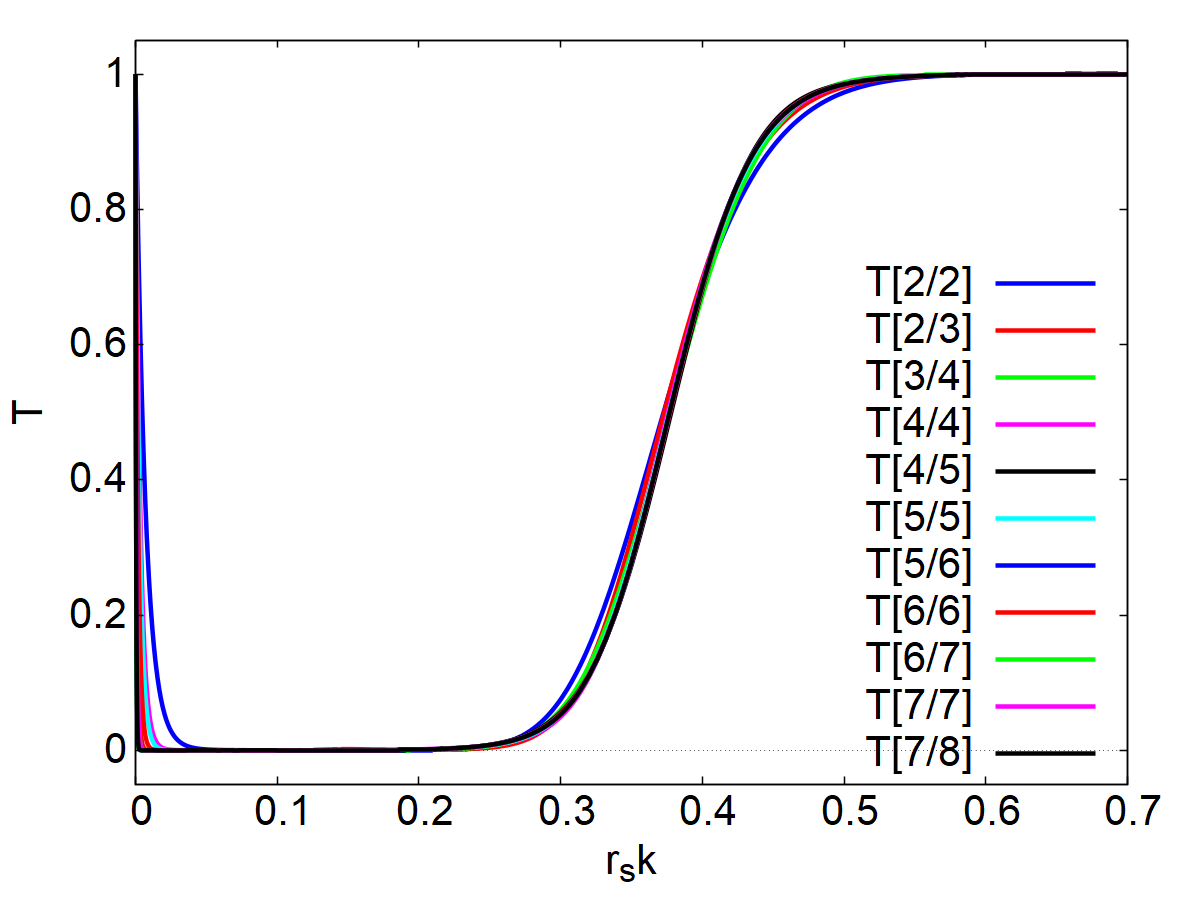}
$\ $
\includegraphics[width=8cm,height=6.5cm]{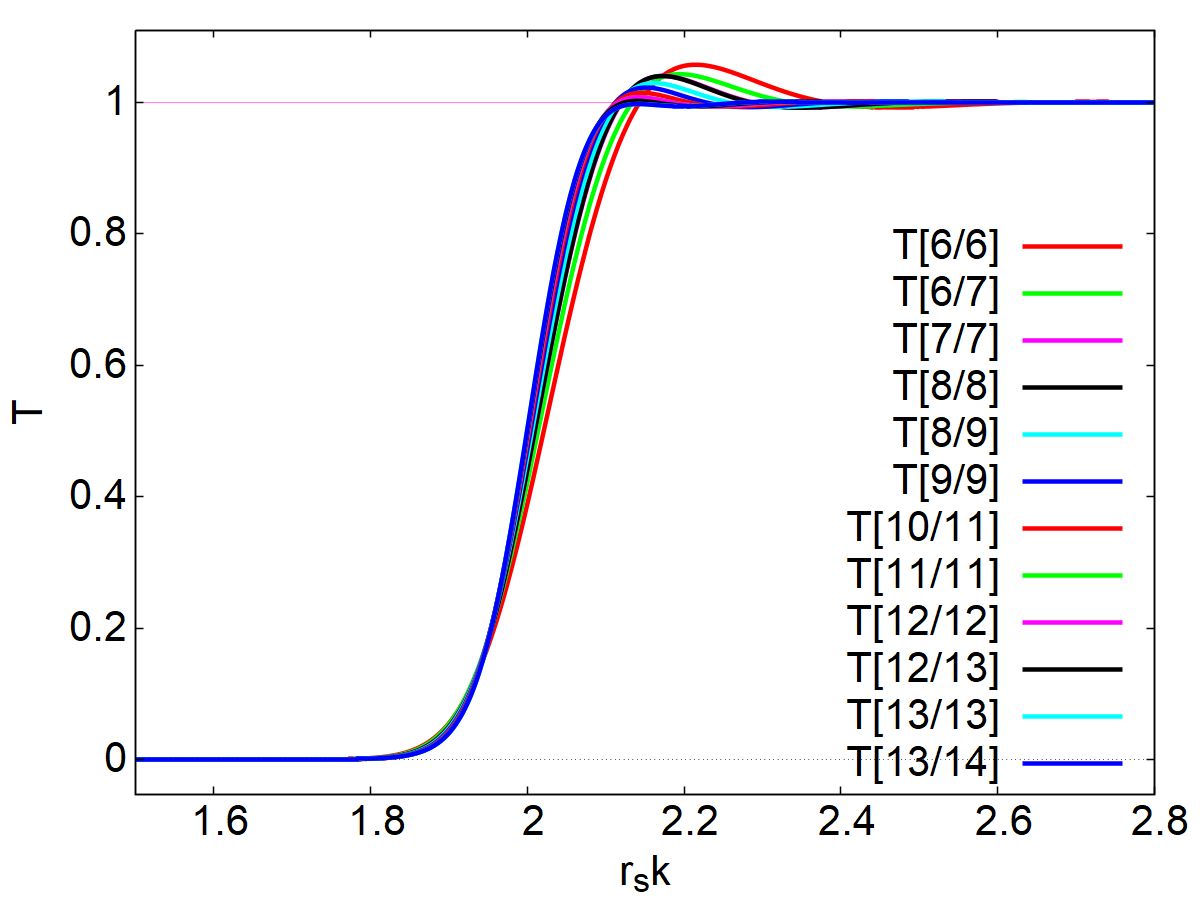}
\caption{Plots of the greybody factors for the Regge-Wheeler potential barrier calculated using the Pad\'e approximation of Eq.~\eqref{greybody-pade} for $\ell=2$ (top panel) and $\ell=10$ (bottom panel). The notation $T[N/M]$ refers to the fact that the greybody factor is calculated from the $[N/M]$ Pad\'e approximant to the MGF asymptotic series of Eq.~\eqref{MGF-asymptotic}. 
\label{Tpade-schw}}
\end{figure}

Returning to our procedure to compute the BH greybody factors, the first step is to find the KdV integrals. To that end, due to the Darboux invariance of the KdV integrals (see Refs.~\cite{Lenzi:2021njy,Lenzi:2022wjv} for details on this symmetry of the space of master functions and equations), we can use any of the potentials that appear in the master equations that describe the perturbations of a Schwarzschild BH, for instance, the Regge-Wheeler potential~\eqref{RW-potential}. The first ten non-vanishing KdV integrals for this potential can be found in Appendix~E of Ref.~\cite{Lenzi:2022wjv}.
The second step is to build the asymptotic series for the MGF [Eq.~\eqref{MGF}]. For this we need the moments, which are proportional to the KdV integrals according to Eq.~\eqref{moments-hamburger}.  
The third step is to construct the Pad\'e approximant to the MGF series to the desired degree of precision. The order of the Pad\'e approximation, $(K,L)$ [see Eq.~\eqref{greybody-pade}] is directly connected to the number of KdV integrals (moments), $n = K+L$, that we use in the computation. 
Then, the fourth step consists in finding the poles and residues of the Pad\'e approximant. This is done numerically and in our case we use the computer algebra system {\it Maxima}~\cite{maxima}. This leads to the semi-analytical approximation of Eq.~\eqref{partial-fraction-form-pade}, in terms of partial fractions, for the Pad\'e approximant. 
The last step is to perform the Laplace inversion, as in Eq.~\eqref{inverse-laplace-hamburger}, to finally obtain the result represented by Eq.~\eqref{greybody-pade}. In Fig.~\ref{Tpade-schw} we show the results for the transmission coefficient using different Pad\'e approximants and for two values of the harmonic number $\ell$: $\ell=2$ (top panel) and $\ell=10$ (bottom panel).

The behavior of the Pad\'e approximation for the greybody factors shown in Fig.~\ref{Tpade-schw} appears to be similar to what we found for the P\"oschl-Teller potential in the previous section. Actually, we observe that the higher the harmonic number $\ell$ (angular momentum), and hence the higher the strength of the potential is, the higher is the number of moments that we need to include in order to obtain a good accuracy for intermediate frequencies (intermediate values of $k$).  

\begin{figure}[t]
\centering
\includegraphics[width=8cm,height=6.5cm]{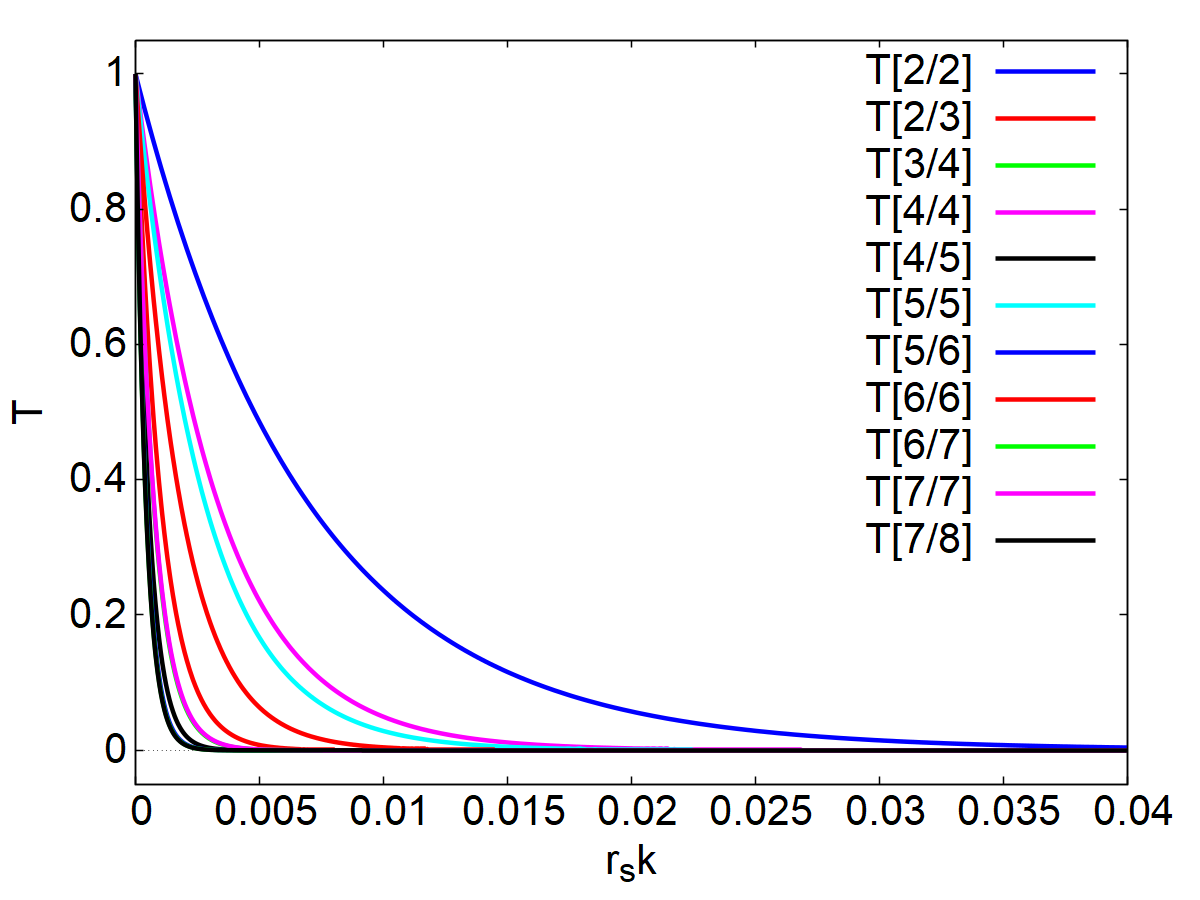}
$\ $
\includegraphics[width=8cm,height=6.5cm]{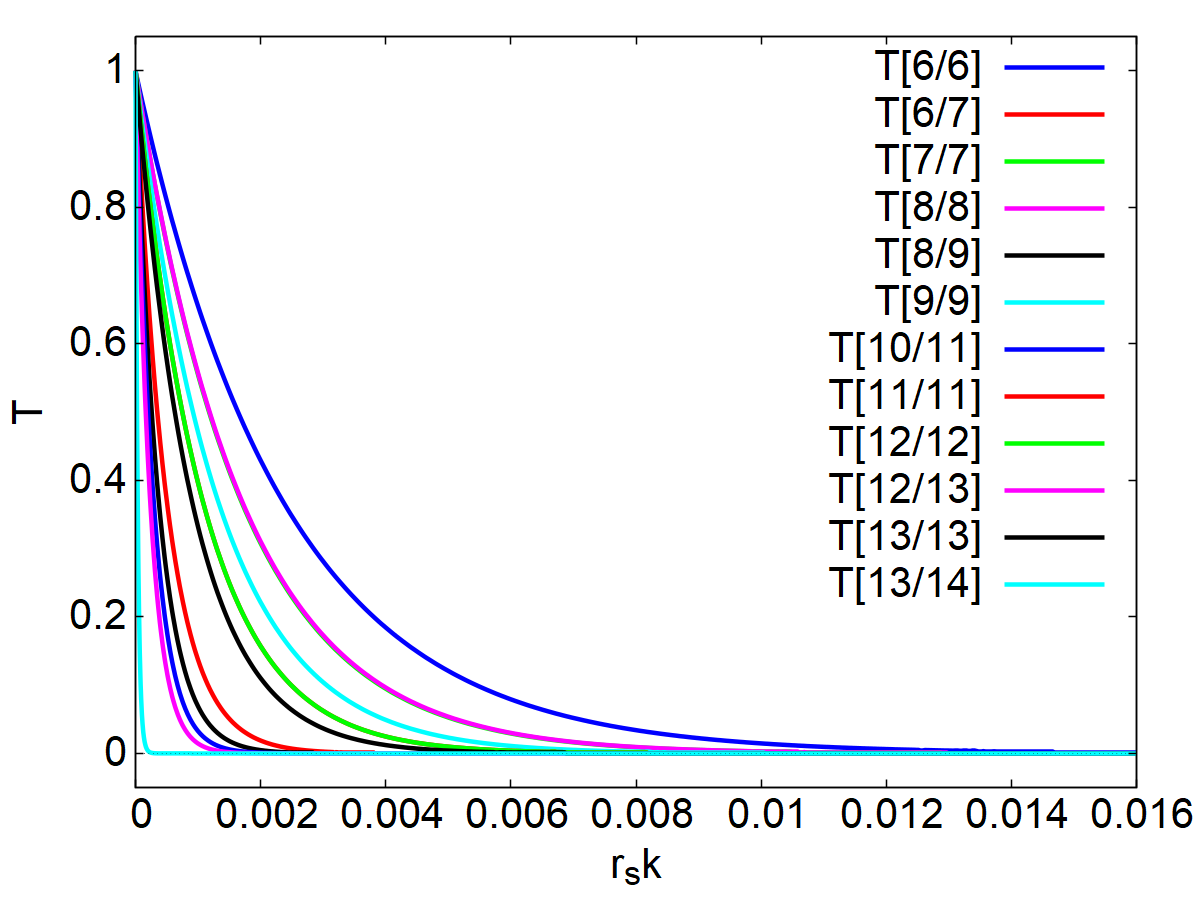}
\caption{The plot shows the behavior of the Regge-Wheeler potential greybody factor in the low-frequency (low $k$) regions obtained by zooming near $k=0$ in plots of Fig.~\ref{Tpade-schw} for $\ell=2$ (top panel) and $\ell=10$ (bottom panel).
\label{Tpade-schw-zoom}}
\end{figure}

\begin{figure}[t]
\centering
\includegraphics[width=8cm,height=6.5cm]{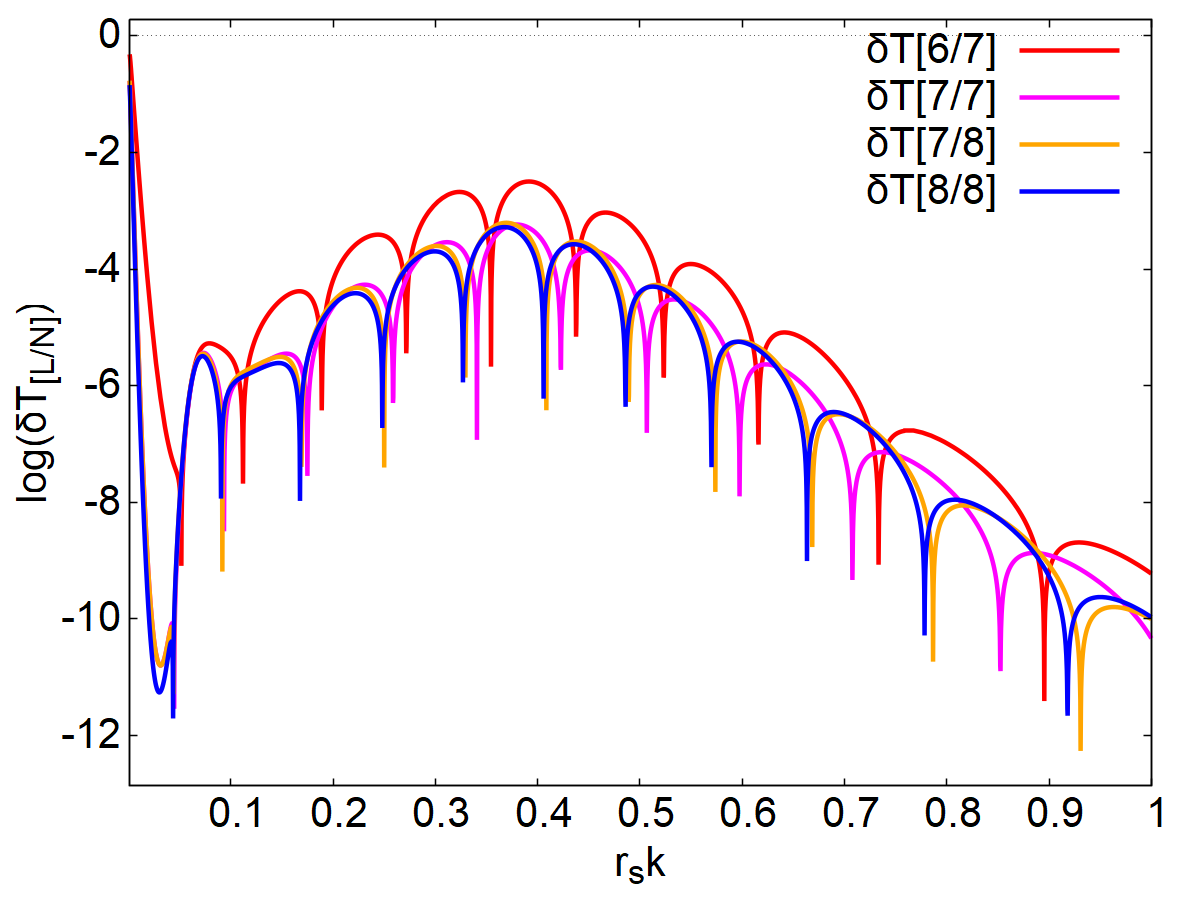}
$\ $
\includegraphics[width=8cm,height=6.5cm]{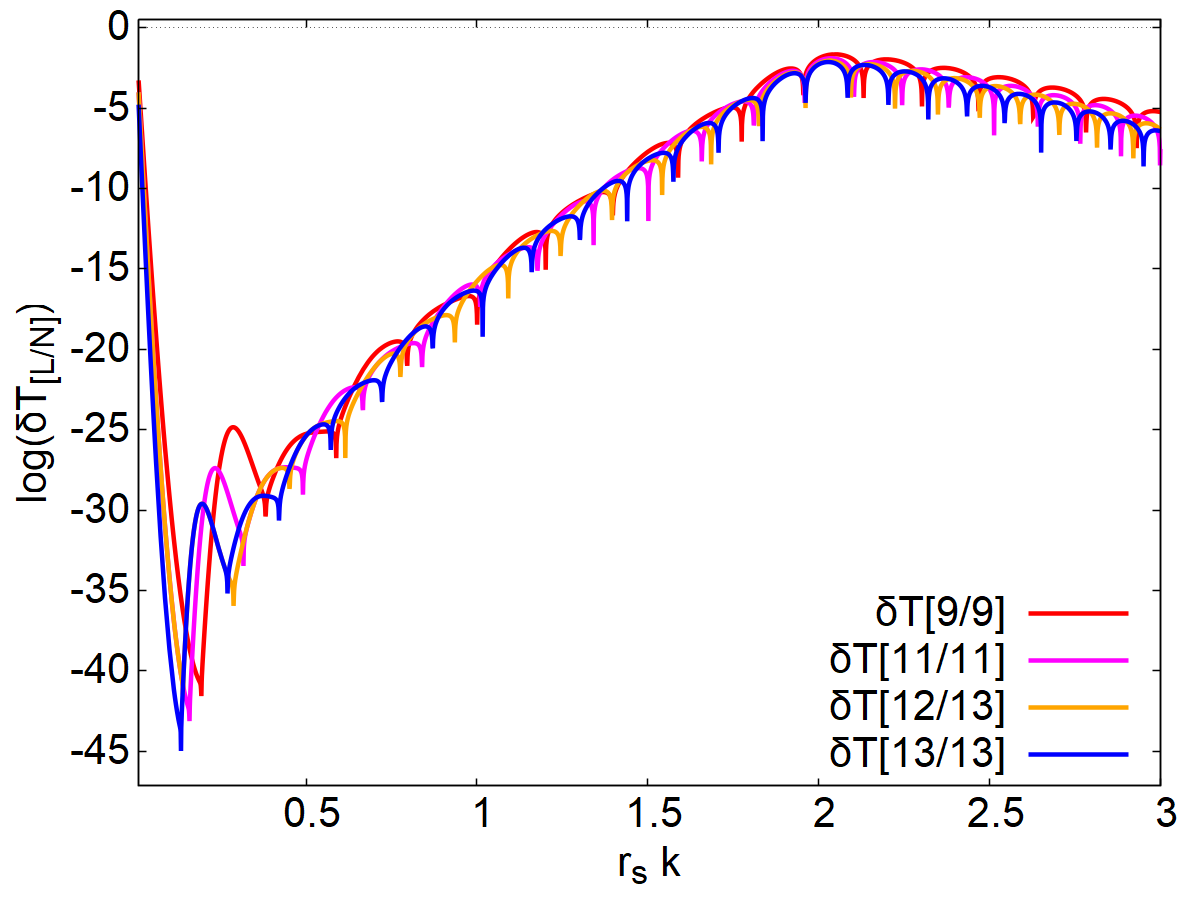}
\caption{Plot of the Regge-Wheeler greybody factor error $\delta T_{[K/L]}(k)$ [Eq.~\eqref{succ-error-T}] for various orders of the Pad\'e approximant used, or equivalently, for different numbers of the KdV integrals involved in the computation. The top panel shows plots for $\ell =2$ and  the bottom panel for $\ell = 10\,$. 
\label{delta-T-RW}}
\end{figure}

On the other hand, in Fig.~\ref{Tpade-schw-zoom} we show the behavior of the Regge-Wheeler greybody factor in the low-frequency (low $k$) region (by zooming into the plots of Fig.~\ref{Tpade-schw}), and in Fig.~\ref{delta-T-RW} we plot the error $\delta T_{[K/L]}$ for successive Pad\'e-based approximations, introduced in Eq.~\eqref{succ-error-T}. By looking at these plots, we can notice again that the higher the harmonic number $\ell$ is, the better the approximation near the origin becomes.  We can also see that the general behavior of the approximation of Eq.~\ref{greybody-pade} is that it improves as we increase the number of moments involved in the computation, as expected. Therefore, following the P\"oschl-Teller case as a guidance, our approximation seems to be very good as we can see by looking at the order of magnitude of $\delta T_{[K/L]}$ in Fig.~\ref{delta-T-RW}

\begin{figure}[t]
\centering
\includegraphics[width=8cm,height=6.5cm]{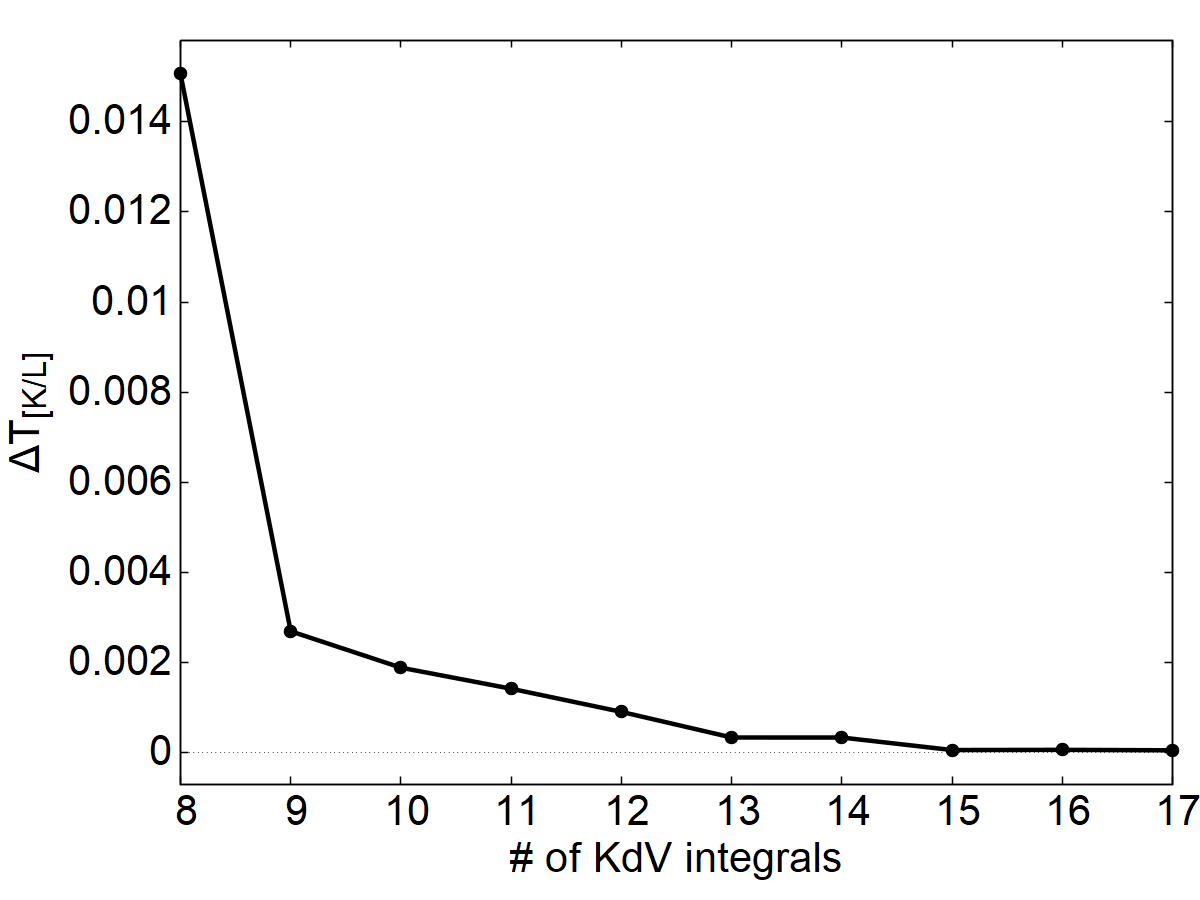}
$\ $
\includegraphics[width=8cm,height=6.5cm]{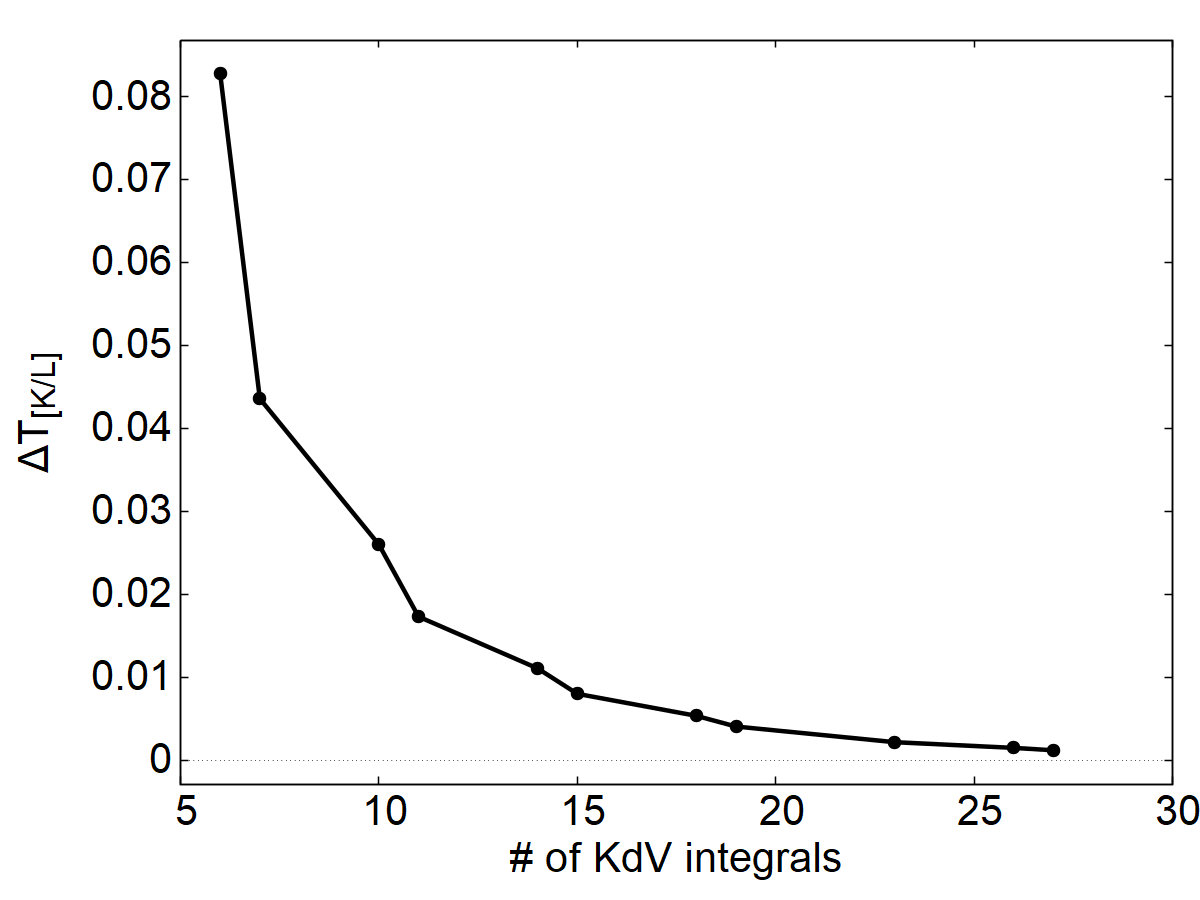}
\caption{Plot of the Regge-Wheeler greybody factor global error $\Delta T_{[K/L]}$ [Eq.~\eqref{abs-error-T}] for various orders of the Pad\'e approximant used, or equivalently, for different numbers of the KdV integrals involved in the computation. The top panel shows plots for $\ell =2$ and cutoffs $(k_0\,,\,k_{\infty}) = (0.05\,,\,20)$, while the bottom panel shows plots for $\ell = 10$ and cutoffs $(k_0\,,\,k_{\infty})= (0.2\,,\,20)\,$. 
\label{DELTA-T-RW}} 
\end{figure}

In Fig.~\ref{DELTA-T-RW} we show two plots (for two values of the harmonic number $\ell = 2, 10$) for the global error estimate~\eqref{abs-error-T} for the Regge-Wheeler greybody factor. Using this error estimate we have a unique measure of the improvement of the approximation with the increase of KdV integrals/moments. Fig.~\ref{DELTA-T-RW} shows the improvement in the approximation as we increase the order of the Pad\'e approximants used in our computations.

In Fig.~\ref{Tpade-Twkb-schw} we plot our Pad\'e-based approximation against the fourth-order WKB approximation (see Appendix~\ref{App:WKB} for details). The comparison shows that the two approximations become almost indistinguishable for high $\ell$ but differ more significantly for lower values of $\ell$. Fig.~\ref{Tpade-Twkb-schw} also shows that by considering more moments (KdV integrals) in the construction of the Pad\'e-based approximation~\eqref{greybody-pade} we can reach very good accuracy (see also Fig.~\ref{Tpade-schw}).

To summarize, we have found that the Pad\'e-based approximation we have introduced in Eq.~\eqref{greybody-pade} provides accurate estimates for the greybody factors associated with gravitational scattering processes around a Schwarzschild BH. The accuracy of the greybody factors computed in this way is comparable to the one obtained from the WKB approximation scheme, provided we use sufficient KdV integrals. In this sense, it is important to mention that while increasing the WKB order of approximation implies a rapid growth of the number of terms that appear at each order~\cite{Matyjasek:2017psv} (see also Appendix~\ref{App:WKB}), the increase in the Pad\'e order (or in other words, in the number of KdV integrals/moments used in the computation) appears to be quite simple and requires a relative quite low increase in computational cost. Therefore, our Pad\'e-based approximation can be pushed to very high-order without neither modifying the procedure nor increasing significantly the computational cost associated with it.

\begin{figure}[t]
\centering
\includegraphics[width=8cm,height=6.5cm]{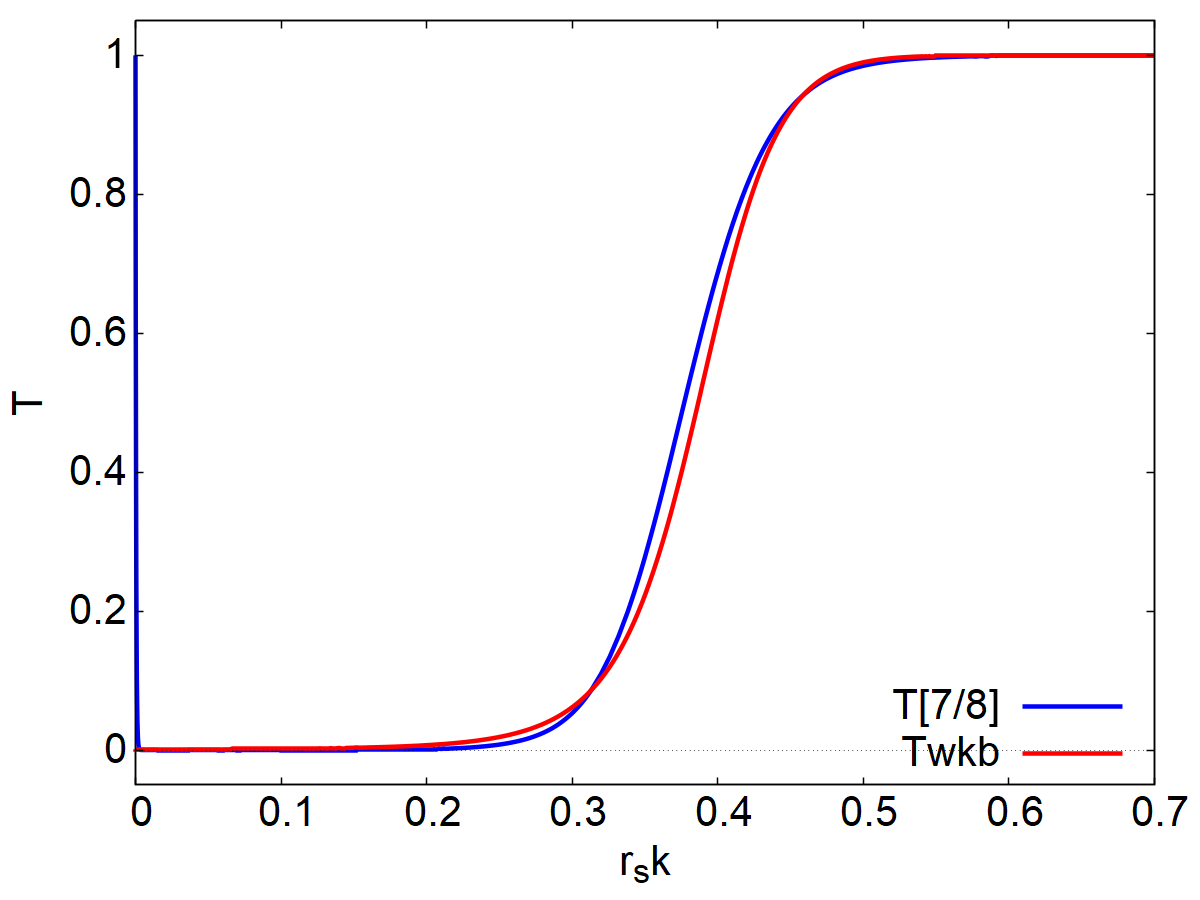}
$\ $
\includegraphics[width=8cm,height=6.5cm]{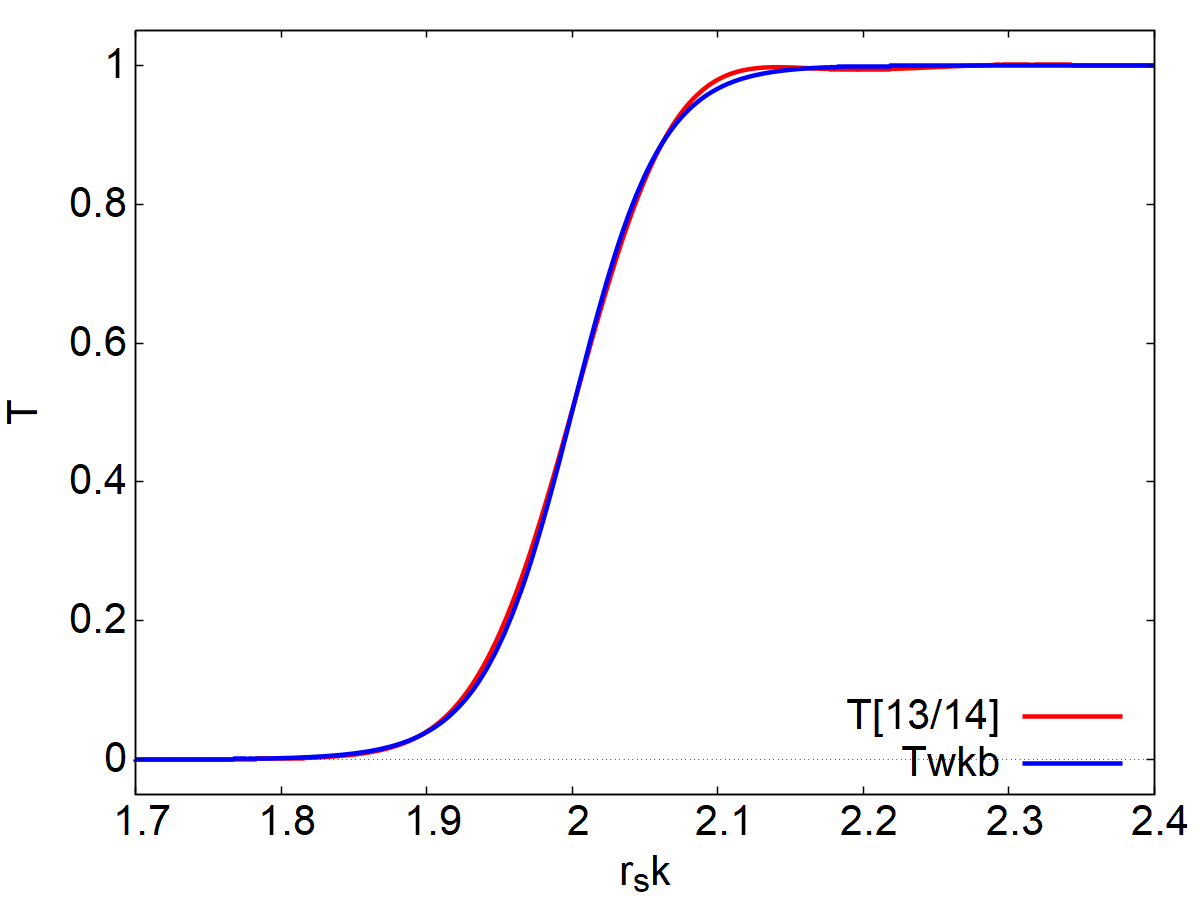}
\caption{Comparison between the Regge-Wheeler greybody factors as computed from the Pad\'e-based approximation with those computed using the fourth-order WKB approximation scheme. The top planel shows the $\ell=2$ case while the bottom panel shows the $\ell=10$ case. 
\label{Tpade-Twkb-schw}}
\end{figure}

\section{Conclusions and Discussion}\label{Sec:Conclusions-and-Discussion}

In this paper we have introduced a new method to compute the greybody factors of a potential barrier, in particular of the gravitational Schwarzschild potential barrier. The method is based on the main result of Ref.~\cite{Lenzi:2022wjv}, which tells us that the greybody factors are uniquely determined by the KdV integrals associated with the BH potential barrier via a moment problem, what we called the 
{\em BH moment problem}.  This result originates from the study of the space of all the possible master functions and equations describing BH perturbations that was carried out in Refs.~\cite{Lenzi:2021njy,Lenzi:2021wpc}. In those works it was found that all the master equations are connected by Darboux Transformations, which preserve the continuum spectrum (the greybody factors) and the quasinormal spectrum (the resonances), showing their physical equivalence.  An additional set of symmetries was found, consisting in deformations that follow the hierarchy of KdV equations and generate an infinite sequence of conservation laws with their associated conserved quantities, the KdV integrals (they are integrals of differential polynomials of the potential that characterize the master equation). It turns out that the KdV integrals are the same for all the possible potentials (for all the possible master equations), that is, they are invariant under Darboux Transformations. In this sense, the KdV integrals characterize the physics behind the master equations, as we have seen with the case of greybody factors and their connection with the KdV integrals via a moment problem. 

There is not a unique way of solving/inverting the moment problem to obtain the greybody factors once the KdV integrals are given. In this work, we have made the choice of using Pad\'e approximants mainly because of their intimate connection to the moment problem, which goes back to the first studies on continued fractions~\cite{AFST_1894_1_8_4_J1_0}. Actually, an important results in this line is that the Pad\'e approximants of a Stieltjes series converge to the corresponding exact Stieltjes function when the problem is determinate~\cite{Bender:1978bo,baker_graves-morris_1996}, i.e. it admits a unique solution (see also~\cite{Lenzi:2022wjv}). Moreover, the Pad\'e approximants have been also proposed as a tool to carry out  the Laplace inversion~\cite{LONGMAN1972224,cohen2007numerical}, so that the moment problem can be solved by inverting the MGF~\cite{Amindavar1994PadeAO}, which is in itself a Laplace transform, to find the distribution function associated with the moments. As a final remark on the use of Pad\'e approximants, it is important to mention that they have been widely used and investigated in the literature and, as a result of this, a number of useful results are known. For instance, the use of Shanks transformations, which improves their convergence rate~\cite{Bender:1978bo}, and/or the multipoint Pad\'e approximants (see, e.g.~\cite{baker_graves-morris_1996}) to improve the approximation near zero frequency.

The results of this paper have to be considered as a first step towards solving the BH moment problem to find the BH greybody factors in this new approach to the problem, introduced in Ref.~\cite{Lenzi:2022wjv}. Despite the good precision we have achieved in the computation of the transmission coefficient, there is still significant room for improvement in the way we use Pad\'e approximants, or even for introducing more sophisticated techniques, which may trigger the exploration of better methods and techniques to solve the BH moment problem. For instance, one possibility would be to look for alternative methods to invert the Laplace transform~\cite{cohen2007numerical,baker_graves-morris_1996}. Leaving aside the approach we use in this paper of introducing the MGF as a main object in the computation, a number of numerical or semi-analytical methods have been studied to invert the moment problem in other contexts (see, for instance~\cite{HULBURT1964555,Marchisio, GAVRILIADIS20124193,Meadpapanicolau, TAGLIANI1999291,JOHN20072890,lebaz:hal-01876363}).

Finally, in this paper, we have applied our techniques to the case of scattering of gravitational perturbations off the potential barrier of a non-rotating BH.  By looking at the details of the whole procedure, it seems reasonable to think that with little modifications, the BH moment problem can be adapted to other physical situations of interest, in particular to the case of BH perturbations of different character (spin): scalar, vector and neutrino perturbations.  Other relevant scenarios where it would be worth applying these techniques include: Perturbations of spinning BHs in General Relativity; perturbations of compact objects, in particular of those exotic cases that can mimic BHs; perturbations of BHs in alternative theories of gravity; perturbations of BHs in higher dimensions; etc.

\begin{acknowledgments}
ML and CFS  are supported by contract PID2019-106515GB-I00/AEI/10.13039/501100011033 (Spanish Ministry of Science and Innovation) and 2017-SGR-1469 (AGAUR, Generalitat de Catalunya). 
ML is also supported by Juan de la Cierva contract FJC2021-047289-I funded by program MCIN/AEI/10.13039/501100011033 (Spanish Ministry of Science and Innovation) and by NextGenerationEU/PRTR (European Union).
This work was also partially supported by the program {\em Unidad de Excelencia Mar\'{\i}a de Maeztu} CEX2020-001058-M (Spanish Ministry of Science and Innovation).
We have used the Computer Algebra System Maxima~\cite{maxima} to carry out most of the computations necessary for this paper. 
\end{acknowledgments}

\appendix

\section{Basics of Pad\'e Approximants} \label{App:Pade-approximants}

Pad\'e approximants (see Ref.~\cite{baker_graves-morris_1996}) are a powerful way of reproducing a power series, in a such a way that the Pad\'e approximant exhibits in general better convergence properties. Let us consider the following power series:
\begin{equation}
f(z) = \sum_{i=0}^{\infty} c^{}_{i} z^{i} \,.
\label{power-series}
\end{equation}
The Pad\'e approximants are a sequence of rational functions of the form
\begin{equation}
 \left[K / L \right](z) 
 =
 \frac{ \sum_{i=0}^{K} P^{}_i z^i}{ \sum_{j=0}^{L} Q^{}_j z^j} \,,
\label{pade-definition}
\end{equation}
such that each term of the sequence is equal to the power series expansion up to the order $K+L+1$, where the $(P_i,Q_j)$ are just constant coefficients. The coefficient $Q_0$ in particular can be fixed to one without loosing generality. The rest of the coefficients are found by matching the expansion of $\left[K / L \right](z) $ up to order $K+L+1$ with the first $K+L+1$ coefficients of Eq.~\eqref{power-series}, i..e.
\begin{equation}
\left(\sum_{i=0}^{K+L} c^{}_{i} z^{i}\right) \left(\sum_{j=0}^{L} Q^{}_j z^j\right) - \sum_{i=0}^{K} P^{}_i z^i = O\left(z^{K+L+1}\right)
\,.
\end{equation}
Then, the coefficients in the denominator of Eq.~\eqref{pade-definition} are related to the coefficients of the power series in the following way
\begin{equation}
\sum_{j=1}^{L}C^{}_{i j} Q^{}_j = c^{}_{K+i}\,,\quad 1\leq j\leq L
\,,
\label{den-pade}
\end{equation}
where $C_{i j} = c_{K +i-j}$ are the element of a $L\times L$ matrix. The coefficients in the numerator are instead found as follows
\begin{equation}
P^{}_n = \sum_{j=0}^{n} c^{}_{n-j} Q^{}_j \,,\quad 0\leq j\leq K
\,,
\label{num-pade}
\end{equation}
where $Q_j=0$ if $j>L$.

\section{WKB Approximations for BH Greybody Factors} \label{App:WKB}

The application of the WKB approximation to the study of perturbations of BHs~\cite{Schutz:1985km} provides useful and accurate expressions for the estimation of quasinormal mode frequencies and greybody factors. In the particular case of greybody factors, the result provided by the WKB approximation can be written as follows~\cite{Iyer:1986np}:
\begin{equation}
T(k) = \frac{1}{\left|1+e^{2\pi i\left(\nu(k) +1/2\right)} \right|}
\,,
\label{greybody-WKB}
\end{equation}
where
\begin{equation}
\nu(k) +1/2
=
i\frac{\left(k^2 -V^{}_0\right)}{\sqrt{-2 V_{0}^{''}}}
-
\sum_{i=2} \Lambda^{}_i \,,
\end{equation}
and
\begin{equation}
V^{}_0 = V(x^{}_0) \,, \quad
V^{(n)}_0 = \left. \frac{d^n V}{dx^n}\right|^{}_{x=x^{}_0} \,,
\end{equation}
where $x_0$ is the location of the maximum of the potential barrier $V(x)$, that is, $x_0$ is defined by the following relation
\begin{equation}
V'(x^{}_0) = 0\,.    
\end{equation}
The quantities $\Lambda_i$ for $i=1,\ldots,13$ can be found in Refs.~\cite{Schutz:1985km,Iyer:1986np,Iyer:1986nq,Konoplya:2003ii,Matyjasek:2017psv}. In particular, the quantities $\Lambda_2\,$, $\Lambda_3\,$, and $\Lambda_4$, the ones used in this work, are given by
\begin{widetext}
\begin{eqnarray}
\Lambda^{}_2 &=& \frac{1}{\sqrt{-2 V_{0}^{''}}} \left[\frac{1}{8}\left(\frac{V_{0}^{(4)}}{V_{0}^{''}} \right)\left( \frac{1}{4}+ \gamma^2 \right) - \frac{1}{288}\left(\frac{V_{0}^{(3)}}{V_{0}^{''}} \right)^2 \left( 7+ 60\gamma^2 \right) \right] \,,
\\
\Lambda^{}_3 &=& \frac{\gamma}{2 V_{0}^{''}}\bigg[\frac{5}{6912}\left( \frac{V_{0}^{(3)}}{V_{0}^{''}}\right)^4 \left( 77+ 188\gamma^2 \right)
-
\frac{1}{384}\left( \frac{V_{0}^{(3)\,2}V_{0}^{(4)}}{V_{0}^{''\,3}}\right) \left( 51+ 100\gamma^2 \right)  
+
\frac{1}{2304}\left( \frac{V_{0}^{(4)}}{V_{0}^{''}}\right)^2 \left( 67+ 68\gamma^2 \right) 
\nonumber \\  
&+&
\frac{1}{288}\left( \frac{V_{0}^{(3)}V_{0}^{(5)}}{V_{0}^{''\,2}}\right) \left( 19+ 28\gamma^2 \right)
-
\frac{1}{288}\left( \frac{V_{0}^{(6)}}{V_{0}^{''}}\right) \left( 5+ 4\gamma^2 \right)
\bigg] \,,
\\ 
\Lambda^{}_4 & = & \frac{1}{597196800 \sqrt{2}V_{0}^{''\,7} \sqrt{V_{0}^{''}}}
\left\{2536975 V_{0}^{(3)\,6}- 9886275 V_{0}^{''}V_{0}^{(3)\,4}V_{0}^{(4)} 
+ 5319720 V_{0}^{''\,2}V_{0}^{(3)\,3}V_{0}^{(5)} -225 V_{0}^{''\,2}V_{0}^{(3)\,2} \right.
\times \nonumber
\\
&\times&
\left(-40261 V_{0}^{(4)\,2} + 9688V_{0}^{''}V_{0}^{(6)}\right) +3240V_{0}^{''\,3}V_{0}^{(3)} \left(-1889 V_{0}^{(4)}V_{0}^{(5)} +220 V_{0}^{''}V_{0}^{(7)} \right) -729 V_{0}^{''\,3} \left[1425 V_{0}^{(4)\,3}  \right.  \nonumber
\\
 &-& \left.\left. 1400 V_{0}^{''}V_{0}^{(4)}V_{0}^{(6)} + 8V_{0}^{''} \left(-123 V_{0}^{(5)\,2} +25 V_{0}^{''}V_{0}^{(8)}  \right) \right] \right\} \nonumber
\\
&+&\frac{\gamma^2}{4976640 \sqrt{2}V_{0}^{''\,7} \sqrt{V_{0}^{''}}}
\left\{ 348425V_{0}^{(3)\,6} - 1199925 V_{0}^{''}V_{0}^{(3)\,4}V_{0}^{(4)} + 57276  V_{0}^{''\,2}V_{0}^{(3)\,3}V_{0}^{(5)} -45V_{0}^{''\,2}V_{0}^{(3)\,2}
\times \right. \nonumber
\\
&\times& \left(-20671 V_{0}^{(4)\,2}+ 4552 V_{0}^{''}V_{0}^{(6)} \right)
+1080 V_{0}^{''\,3}V_{0}^{(3)} \left( -489 V_{0}^{(4)}V_{0}^{(5)} +52 V_{0}^{''} V_{0}^{(7)}  \right) -27 V_{0}^{''\,3}\left[2845V_{0}^{(4)\,3} \right. \nonumber
\\
&-& \left. \left.2360 V_{0}^{''}V_{0}^{(4)}V_{0}^{(6)} + 56 V_{0}^{''} \left( -31V_{0}^{(5)\,2} + 5V_{0}^{(2)}V_{0}^{(8)}\right)\right]
\right\} \nonumber
\\
&+&
\frac{\gamma^4}{2488320 \sqrt{2}V_{0}^{''\,7} \sqrt{V_{0}^{''}}}
\left\{ 192925 V_{0}^{(3)\,6}  - 581625 V_{0}^{''}V_{0}^{(3)\,4}V_{0}^{(4)} + 234360 V_{0}^{''\,2}V_{0}^{(3)\,3}V_{0}^{(5)}  -45V_{0}^{''\,2}V_{0}^{(3)\,2}
\times \right. \nonumber
\\
&\times& \left(-8315 V_{0}^{(4)\,2}+ 1448 V_{0}^{''}V_{0}^{(6)} \right) +1080 V_{0}^{''\,3}V_{0}^{(3)} \left( -161 V_{0}^{(4)}V_{0}^{(5)} +12 V_{0}^{''} V_{0}^{(7)}  \right) -27 V_{0}^{''\,3}\left[625 V_{0}^{(4)\,3} \right. \nonumber
\\
&-& \left. \left. 440 V_{0}^{''}V_{0}^{(4)}V_{0}^{(6)} + 8 V_{0}^{''} \left( -63V_{0}^{(5)\,2} + 5V_{0}^{(2)}V_{0}^{(8)}\right)\right]
\right\}
\,.
\end{eqnarray}
\end{widetext}
where $\gamma = n +1/2$ and $n = 0,1,2...$ is the overtone number.

\section{Some Formulas for the P\"oschl-Teller Potential} \label{App:PT}
Let us consider the time-independent Schr\"odinger equation~\eqref{schrodinger} for the case of a P\"oschl-Teller potential barrier [see Eq.~\eqref{pt-potential-barrier}. The general solution reads (Refs.~\cite{landau1981quantum,Cevik:2016mnr, Lenzi:2022wjv}):
\begin{widetext}
\begin{eqnarray}
\nonumber
\psi(x,k)
&=&
A\, 2^{ik/\alpha} \left[1 - \tanh^2(\alpha x)\right]^{-ik/2\alpha}
{}_{2}F_1 \left(\lambda -\frac{i k}{\alpha}, 1- \lambda -\frac{i k}{\alpha}  , 1-\frac{i k}{\alpha};\frac{1 - \tanh(\alpha x)}{2} \right)
\\
&+&
B\, \left[1 - \tanh(\alpha x)\right]^{ik/2\alpha} 
\left[1 + \tanh(\alpha x)\right]^{-ik/2\alpha}
{}_{2}F_1 \left(\lambda, 1- \lambda , 1+\frac{i k}{\alpha};\frac{1 - \tanh(\alpha x)}{2} \right)
\,,
\end{eqnarray}
\end{widetext}
where
\begin{equation}
\lambda = i\beta + \frac{1}{2} \,.    
\end{equation}
By analyzing the asymptotic behavior of the general solution, both at $x \rightarrow \infty $ and at $x \rightarrow -\infty$, we can find analytical
expressions for the reflection and transmission coefficients. First, one can obtain the following expressions for the Bogoliubov coefficients
\begin{eqnarray}
a(k) & = & \frac{\Gamma\left(1-ik\right) \Gamma\left(-ik\right)}{\Gamma\left(\frac{1}{2}-i(k-\beta)\right)\Gamma\left(\frac{1}{2}-i(k+\beta)\right)} \,,
\label{ak-pt}
\\
b(k) & = & \frac{\Gamma\left(1-ik\right) \Gamma\left(ik\right)}{\Gamma\left(\frac{1}{2}+i\beta)\Gamma(\frac{1}{2}-i\beta\right)} \,,
\end{eqnarray}
where for the sake of simplicity we set $\alpha =1$. Then, we can find the transmission and reflection coefficients from Eq.~\eqref{reflection-transmission-coefficient}. 
The transmission probability can be evaluated by taking the inverse modulus square of Eq.~\eqref{ak-pt} for real $k$ and reads
\begin{eqnarray}
T(k) =
\frac{\sinh^2\left(\pi k\right)}{\cosh^2\left(\pi k\right)+\sinh^2\left(\pi \beta\right)}
\,,
\label{pt-transmission-real-k}
\end{eqnarray}
where we used the properties of the gamma functions~\cite{Abramowitz:1970as}.

\section{KdV Integrals for the P\"oschl-Teller Potential} \label{App:KdV-PT}

Here we list the first non-vanishing KdV integrals [see Eq.~\eqref{kdv-integrals}] for the P\"oschl-Teller potential [see Eq.~\eqref{pt-potential-barrier}]:
\begin{widetext}
\begin{eqnarray}
I^{}_1 & = & \frac{\alpha}{2} \left(4 \beta^{2} + 1 \right) \,, 
\\
I^{}_3 & = & -\frac{\alpha^3}{12} \left(4 \beta^{2} + 1 \right)^2 \,, 
\\
I^{}_5 & = & \alpha^5 \left(4 \beta^{2} + 1 \right)^2 \left( \frac{2}{15} \beta^2 - \frac{1}{6} \right) \,, 
\\
I^{}_7 & = & -\alpha^7 \left(4 \beta^{2} + 1 \right)^2 \left( \frac{2}{7} \beta^4
- \frac{41}{105}\beta^2 - \frac{577}{840} \right) \,, 
\\
I^{}_{9} & = &  \alpha^9 \left(4 \beta^{2} + 1 \right)^2 \left( \frac{32}{45} \beta^6 - \frac{232}{315} \beta^4 - \frac{1406}{315} \beta^2 -\frac{2683}{630} \right) \,, 
\\
I^{}_{11} & = & - \alpha^{11} \left(4 \beta^{2} + 1 \right)^2 \left(
\frac{64}{33} \beta^8  - \frac{64}{99} \beta^6 - \frac{12968}{693} \beta^4 - \frac{168668}{3465} \beta^2  - \frac{61767}{1540} \right) \,,
\\
I^{}_{13} & = &  \alpha^{13} \left(4 \beta^{2} + 1 \right)^2  \left(
\frac{512}{91} \beta^{10} + \frac{12160}{3003} \beta^8 - \frac{956096}{15015} \beta^6 - \frac{1123952}{3465} \beta^4 - \frac{2879438}{4095} \beta^2 - \frac{49126459}{90090} \right) \,, 
\\
I^{}_{15} & = & -\alpha^{15} \left(4 \beta^{2} + 1 \right)^2  \left(
\frac{256}{15} \beta^{12} + \frac{1408}{39} \beta^{10} - \frac{1057744}{6435} \beta^8 - \frac{826544}{495} \beta^6 - \frac{303716683}{45045} \beta^4  - \frac{406249013}{30030} \beta^2 \right. \nonumber \\
& &  \left. - \frac{813135229}{80080} \right) \,,
\\
I^{}_{17} & = & \alpha^{17} \left(4 \beta^{2} + 1 \right)^2 \left(
\frac{8192}{153} \beta^{14} + \frac{157696}{765} \beta^{12} - \frac{436736}{3315} \beta^{10} - \frac{755645056}{109395} \beta^8 - \frac{116271200}{2431} \beta^6 - \frac{134553980488}{765765} \beta^4 \right. \nonumber \\
& & \left. - \frac{51719984938}{153153} \beta^2 - \frac{126841684787}{510510} \right) \,,
\\
I^{}_{19} & = & -\alpha^{19} \left(4 \beta^{2} + 1 \right)^2  \left(
\frac{16384}{95} \beta^{16} + \frac{4947968}{4845} \beta^{14} + \frac{10694656}{4845} \beta^{12} - \frac{2947115008}{146965} \beta^{10} - \frac{119163316352}{440895} \beta^8 \right. \nonumber \\
& & \left. - \frac{720103571584}{440895} \beta^6 
 - \frac{115858335632}{20349} \beta^4 + \frac{14090850111224}{1322685} \beta^2 - \frac{8185650267425}{1058148} \right) \,.
\end{eqnarray}
\end{widetext}
%

%

\end{document}